\documentclass[12pt]{article}
\usepackage{graphicx,amsmath,amssymb,booktabs,enumerate}
\usepackage{subfigure, url}
\usepackage{comment}
\usepackage{color}

\def\a{\alpha}
\def\b{\beta}

\def\m{\mu}
\def\n{\nu}

\def\s{\sigma}
\def\t{\tau}

\def\G{\Gamma}

\def\beq{\begin{eqnarray}}
\def\eeq{\end{eqnarray}}
\newcommand{\gsim}{ \mathop{}_{\textstyle \sim}^{\textstyle >} }
\newcommand{\lsim}{ \mathop{}_{\textstyle \sim}^{\textstyle <} }

\newcommand{\GEV}{ {\rm GeV} }
\newcommand{\TEV}{ {\rm TeV} }

\def\tr{\mathop{\rm tr}}

\allowdisplaybreaks[1]

\begin{document}
\baselineskip 0.7cm
\begin{titlepage}

\begin{flushright}
IPMU-13-0143\\
KEK-TH-1641
\end{flushright}

\vskip 1.35cm
\begin{center}
{\large \bf 
Flavor of Gluino Decay in  High-Scale Supersymmetry
}

\vskip 1.2cm
Ryosuke Sato${}^{1}$, Satoshi Shirai${}^{2}$ and Kohsaku Tobioka${}^{3}$
\vskip 0.4cm

{\it
$^1$ Institute of Particle and Nuclear Studies,\\
High Energy Accelerator Research Organization (KEK)\\
Tsukuba 305-0801, Japan\\
$^2$Berkeley Center for Theoretical Physics, Department of Physics, \\
and Theoretical Physics Group, Lawrence Berkeley National Laboratory, \\
University of California, Berkeley, CA 94720, USA \\
$^3$Department of Physics, University of Tokyo, Tokyo 113-0033, Japan,\\
and Kavli Institute for the Physics and Mathematics of the
  Universe (WPI), 
  \\Todai Institutes for Advanced Study, University of Tokyo, \\Kashiwa 277-8583, Japan\\
}

\vskip 1.5cm

\abstract{
The discovery of the Higgs boson with a mass of about 125 GeV intimates us a possibility of a high-scale supersymmetry (SUSY) breaking model, 
where a sfermion mass scale is much higher than the electroweak scale.
Although a general SUSY standard model can contribute to the low-energy flavor and/or CP-violating processes,
the high-scale SUSY breaking model provides smaller signatures and therefore are less constrained, 
even in the presence of large flavor/CP violations of sfermions.
However, a manner of gluino decay directly reflects the squark flavor structure and provides us a clue for the sfermion flavor structure.
In this paper, we study the gluino decay in detail and discuss the interplay with the gluino decay and low-energy flavor and CP observation.

}
\end{center}
\end{titlepage}
\setcounter{page}{2}

%%%%%%%%%%%%%%%%%%%%%%%%%%%%%%%%%%%%%%%%%%%%%%%%%%%%%%%%
% intro
%%%%%%%%%%%%%%%%%%%%%%%%%%%%%%%%%%%%%%%%%%%%%%%%%%%%%%%%
\section{Introduction}
A supersymmetric (SUSY) standard model (SSM) is a leading candidate of physics beyond the standard model (SM).
However the most ``natural'' SUSY model suffers from long-standing problems, such as 
low-energy flavor/CP physics, direct sparticle search, Higgs boson mass, and cosmology.
Many types of SSMs are studied to get over such difficulties.

The discovery of the 125 GeV Higgs boson \cite{Aad:2012tfa, Chatrchyan:2012ufa} indicates a possibility of a very simple framework of SSMs, i.e.,
high-scale SUSY scenarios \cite{Wells:2003tf,ArkaniHamed:2004fb, Giudice:2004tc, ArkaniHamed:2004yi,Wells:2004di}.
The sfermion mass scale higher than ${\cal O}(10)$ TeV can explain for the 125 GeV Higgs boson~\cite{OYY, Giudice:2011cg}.
In addition, constraints of the SUSY flavor/CP problems and cosmological problems are greatly relaxed.
As for the fermion sectors, the R-parity conservation sets gauginos and/or Higgsino mass scale to a TeV range to avoid the dark matter overabundance, 
which is also preferred by the gauge coupling unification.
From a theoretical viewpoint, such a setup can be simply realized. 
For example, an assumption of some charge of the SUSY breaking field $X$ leads the sfermion mass $m_{\rm sfermion} \gsim m_{3/2}$,
keeping the gauginos light, $m_{\rm gaugino} \sim 0.01 m_{3/2}$, which are generated via anomaly mediation effects~\cite{AMSB}.
By setting $m_{3/2}={\cal O}(10^{2-3})$ TeV, a viable SUSY mass spectrum can be obtained.
After the discovery of the Higgs boson, such a framework gets more attention, variously called
 ``Spread SUSY'' \cite{spread}, ``Pure Gravity Mediation'' \cite{PGMs}, ``Mini-Split'' \cite{Arvanitaki:2012ps},  ``Simply Unnatural SUSY'' \cite{ArkaniHamed:2012gw} and so on. 
This framework predicts a Wino is the lightest SUSY particle (LSP).
There are intensive studies of collider \cite{Ibe:2006de, Asai:2007sw} and dark matter \cite{Gherghetta:1999sw,Hisano:2005ec, Hisano:2010ct, Hisano:2004ds, Moroi:2011ab, Cohen:2013ama, Fan:2013faa} signals of the Wino LSP.

Such a large mass hierarchy the between gaugino and scalar sector is not an uncommon feature of SUSY breaking models.
This is because the mass of fermions can be protected from large radiative corrections thanks to some chiral
symmetry and, on the other hand, that of the scalar particle is not protected, once SUSY is broken.
For example, the large mass hierarchy  also exists in some class of gauge mediated SUSY breaking (GMSB) models.
See, e.g., Refs. \cite{Izawa:1997gs, Nomura:1997uu} for such models.
This feature quite often comes from underlying symmetry of the model and/or stability conditions of the SUSY breaking vacua \cite{Komargodski:2009jf, Shirai:2010rr}.
Motivated by the discovery of the Higgs boson, such GMSB models also have been studied \cite{Arvanitaki:2012ps, Ibe:2012dd}.

Several kinds of SUSY breaking models realize the high-scale SUSY spectrum.
Although they predict similar spectra at a TeV scale, 
the gravitino mass and sfermion sector have strong model dependence.
One of the characteristic features of each model appears in the sfermion sector.
For example, without any flavor symmetry, we can expect flavor violating soft parameters of the order of the scalar mass scale.
On the other hand, some flavor symmetry forbid such terms, depending on the symmetry charge assignments.
Or a typical GMSB model predicts just the minimal flavor violation.
A grand unified theory (GUT) may determine the mass relation among the sfermion sector.
Therefore, probing the sfermion sector is essential to study the underlying model.
The information on the sfermion sector is also important for us to discuss the prospects of precise measurements of 
flavor/CP experiments.

For the future flavor/CP observations and deeper understanding the underlying theory of the high-scale SUSY models,
it is very important to know the property of the sfermion sector, such as its mass scale and flavor structure.
Since the direct production of heavy sfermions is out of range of the LHC,
we must consider indirect ways to probe the sfermion sector.
A cosmological signature such as gravitational waves is an interesting possibility among them \cite{Saito:2012bb}.
Another possibility is the gluino decay since it is sensitive to the squark sector.
The most known feature of the gluino in the high-scale SUSY models is its longevity of the lifetime.
Another interesting point is that the flavor violating decay modes of the gluino are allowed
thanks to heavy squark masses satisfying flavor constraints for the SM particles.
Hence, by studying the gluino decay in detail, we can get insights into the squark sectors.
In this paper we study the gluino decay in the high-scale SUSY models.

For heavy squarks, quantum corrections to the gluino decay get large, and can significantly affect the gluino decay, compared to the tree-level calculation.
Previously, there are some works to study the tree-level gluino decay with flavor/CP violation \cite{Toharia:2005gm} and
the quantum effect without flavor/CP violation in the high-scale SUSY models \cite{Gambino:2005eh,Sato:2012xf}.
In this study, we pay particular attention to the quantum effects on the flavor structure of the gluino decay and
discuss the correlation with the low-energy flavor/CP violating rare processes.
In section 2, we estimate the size of the quantum corrections for the gluino decay, using renormalization group equations.
In section 3, we discuss the correlation between the low-energy flavor/CP observations and the gluino decay.
In section 4, we briefly discuss the collider signatures.
Section 5 is devoted to summary and discussions.

%%%%%%%%%%%%%%%%%%%%%%%%%%%%%%%%%%%%%%%%%%%%%%%%%%%%%%%%
% RGE running of Wilson coefficients
%%%%%%%%%%%%%%%%%%%%%%%%%%%%%%%%%%%%%%%%%%%%%%%%%%%%%%%%
%%%%%%%%%%%%%%%%%%%%%%%%%%%%%%%%%%%%%%%%%%%%%%%%%%%%%%%%
% Wilson coeff. and RGE
%%%%%%%%%%%%%%%%%%%%%%%%%%%%%%%%%%%%%%%%%%%%%%%%%%%%%%%%
\section{Gluino Decay in High-scale SUSY Models}
In the high-scale SUSY models, we cannot directly probe the sfermions at the LHC.
However, the gluino decay requires virtual squark exchanges and therefore it reflects the squark sector. 
The most significant feature of the high-scale SUSY model is the longevity of the gluino lifetime 
since, as the squark mass scale goes high, the lifetime of the gluino can be very long.
Typically, its decay length is $c\tau_{\tilde g} \sim 1~{\rm cm} (m_{\tilde q}/1000~{\rm TeV})^4  (m_{\tilde g}/1~{\rm TeV})^{-5}$.
Such a long lifetime plays a significant role not only in collier signals but also in cosmology.
Cosmological constraints disfavor the gluino lifetime much longer than 1 sec \cite{Arvanitaki:2005fa}.

Another important feature is that the decay products of the gluino carry information on details of the squark flavor structure.
For instance, if the third family squarks are lighter than the other squarks, 
the final states of the gluino decay are rich in bottom and top quarks. 
Furthermore, with the flavor violating structure of the squarks, we may observe the flavor violating gluino decay. 
In this way, the squark sector is testable via the gluino decay. 
To extract information on the sfermion sector via observation of the gluino decay, 
we need to sophisticate a prediction of the gluino decay, including quantum corrections.

In this section, we calculate partial decay widths of the gluino by a renormalization group (RG) method.
We study the RG evolution of Wilson coefficients that govern the gluino decay with general flavor/CP structure and
see the effects on the gluino decay.
\subsection{Effective Theory and RG Equations of Wilson Coefficients}
We assume that sfermions, heavy Higgs bosons and Higgsinos have the masses of around $\tilde m > {\cal O}(10)~\TEV$ 
and that the size of $A$-terms is negligible. 
The soft mass terms relevant to our study are
\begin{align}
{\cal L}_{\rm soft} = -\frac{1}{2}\left(m_{\tilde{B}}\tilde{B}\tilde{B} +m_{\tilde{W}}\tilde{W}^A \tilde{W}^A +m_{\tilde{g}}\tilde{g}^a \tilde{g}^a + h.c. \right) 
- {\tilde Q_L}^* m^2_{\tilde Q_L}{\tilde Q_L} - {\tilde u_R}^{c} m^2_{\tilde u_R}{\tilde u_R^{c*}} - {\tilde d_R}^{c} m^2_{\tilde d_R}{\tilde d_R^{c*}},
\end{align}
where $m_{\tilde{B}}$, $m_{\tilde{W}}$ and $m_{\tilde{g}}$ are complex gaugino masses and 
 $m^2_{\tilde Q_L}$,  $m^2_{\tilde u_R}$ and  $m^2_{\tilde d_R}$ are $3\times 3$ Hermitian matrices whose components are ${\cal O}(\tilde m^2)$.
In this section, we take weak basis in which $Q_{L,i}$ and $\tilde Q_{L,i}$ form an $SU(2)_L$ doublet for each $i$.
Thanks to this notation, we can write effective interactions and RG equations for their Wilson coefficients as $SU(2)_L$ symmetric forms. %\footnote{
For a detail, see the Appendix \ref{sec:notation}.
 
Below the scale $\tilde m$, the dynamics of the particles is described by an effective theory
which contains the SM particles and gauginos.
The gluino will mainly decay into a lighter SUSY particle by emitting two quarks or a gluon.
These decay modes are dominantly induced by the following dimension five and six effective interactions by integrating out the heavy squarks:
\begin{eqnarray}
Q_{Q,ij}^{\tilde B} &=& (\tilde B^\dagger \bar\s^\m \tilde g^a) (Q_{L,i}^\dagger \bar\s_\m T^a Q_{L,j} ),\\
Q_{u,ij}^{\tilde B} &=& (\tilde B^\dagger \bar\s^\m \tilde g^a) (u_{R,i}^c \s_\m T^a u_{R,j}^{c\dagger} ),\\
Q_{d,ij}^{\tilde B} &=& (\tilde B^\dagger \bar\s^\m \tilde g^a) (d_{R,i}^c \s_\m T^a d_{R,j}^{c\dagger} ),\\
Q_7^{\tilde B} &=& (\tilde B \bar\s^{\m\n} \tilde g^a) G^a_{\m\n},\\
Q_{ij}^{\tilde W} &=& (\tilde W^{A\dagger} \bar\s^\m \tilde g^a) (Q_{L,i}^\dagger \bar\s_\m T^a \t^A Q_{L,j} ). \label{eq:op_wino}
\end{eqnarray}
Here, $i, j = 1,2,3$ are flavor indices, $A = 1,2,3$ is an $SU(2)_L$ adjoint index and $a= 1,\cdots,8$ is an $SU(3)_C$ adjoint index.

The effective Lagrangian contains the following interaction terms:
\begin{eqnarray}
{\cal L}_{\rm eff.} ~=~ \frac{1}{\tilde m^2} \left[ \sum_{i=1}^3 \sum_{j=1}^3 \left(
C^{\tilde B}_{Q,ij} Q^{\tilde B}_{Q,ij}
+ C^{\tilde B}_{u,ij} Q^{\tilde B}_{u,ij}
+ C^{\tilde B}_{d,ij} Q^{\tilde B}_{d,ij}
+ C^{\tilde W}_{ij} Q^{\tilde W}_{ij}
\right) + C^{\tilde B}_7 Q^{\tilde B}_7 \right] + h.c..
\end{eqnarray}

\subsubsection*{Boundary condition at the squark mass scale}
We have to match our effective theory with the MSSM at the squark mass scale $\tilde m$.
The Wilson coefficients $C(\tilde m)$'s are obtained by integrating out the squarks.
At the leading order, we obtain,
\begin{eqnarray}
C^{\tilde B}_{Q,ij}(\tilde m) &=& -\frac{g_s g'}{6}\tilde m^2 (m^2_{\tilde Q_L})^{-1}_{ij},\\
C^{\tilde B}_{u,ij}(\tilde m) &=& \frac{2g_s g'}{3}\tilde m^2 (m^2_{\tilde u_R})^{-1}_{j i},\\
C^{\tilde B}_{d,ij}(\tilde m) &=& -\frac{g_s g'}{3}\tilde m^2 (m^2_{\tilde d_R})^{-1}_{j i},\\
C^{\tilde B}_7(\tilde m) &=& \frac{g_s^2 g'}{384\pi^2}(m_{\tilde g} - m_{\tilde B}) \tilde m^2 \left(
 {\rm tr} [(m^2_{\tilde Q_L})^{-1}]
- 2 {\rm tr} [(m^2_{\tilde u_R})^{-1}]
+  {\rm tr} [(m^2_{\tilde d_R})^{-1}]
\right), \\
C^{\tilde W}_{ij}(\tilde m) &=& -\frac{g_s g}{2}\tilde m^2 (m^2_{\tilde Q_L})^{-1}_{ij}.
\end{eqnarray} 
Here, $m_{\tilde Q_L}^2$, $m_{\tilde u_R}^2$ and $m_{\tilde d_R}^2$ are the squark mass-squared matrices at the scale $\tilde m$.
We treat the Wilson coefficients $C^{\tilde B}_Q$, $C^{\tilde B}_u$, $C^{\tilde B}_d$ and $C^{\tilde W}$ as $3\times 3$ matrices in flavor space.
In the high-scale SUSY models, there is a large hierarchy between the gaugino mass scale and the squark mass scale $\tilde m$.
Therefore, the resummation of leading logarithm corrections becomes important.

\subsubsection*{RG equations for Wilson coefficients}
\begin{figure}[t!]
\begin{center}
\includegraphics[width=110 mm]{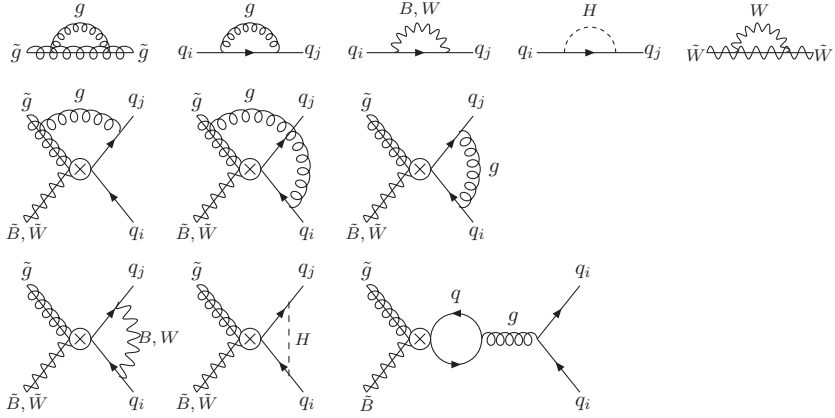}
\caption{ One-loop diagrams for the anomalous dimensions of quarks and gauginos (top)
and the anomalous dimensions of operators and operator mixings (middle and bottom).
}\label{fig:RGdiagram}
\end{center}
\end{figure}

Now, we describe the RG evolution of the Wilson coefficients at one-loop level.
We treat SM-like Yukawa couplings $Y_u$ and $Y_d$ as $3\times 3$ matrices in flavor space. They are defined as,
\begin{eqnarray}
{\cal L}_{\rm yukawa} = H Q_L Y_u u_R^c + H^* Q_L Y_d d_R^c + h.c..
\end{eqnarray}
Here, $H$ is the SM-like Higgs doublet field.
RG equations for the gauge coupling constants are given by,
\begin{eqnarray}
16\pi^2 \frac{d g'}{d\log\m} ~=~ \frac{41}{6} g'^3, ~~~~~
16\pi^2 \frac{d g}{d\log\m} ~=~ -\frac{11}{6} g^3, ~~~~~
16\pi^2 \frac{d g_s}{d\log\m} ~=~ -5 g_s^3. 
\end{eqnarray}
For the quark Yukawa couplings,
\begin{align}
16\pi^2 \frac{d Y_u}{d\log\m}&=\left[  \frac{3}{2} Y_u Y_u^\dagger - \frac{3}{2} Y_d Y_d^\dagger 
+3 {\tr} \left(Y_u Y_u^\dagger + Y_d Y_d^\dagger \right) - \frac{17}{12}g'^2 - \frac{9}{4} g_2^2 - 8 g_s^2 \right] Y_u, \label{eq:RGE_yu}\\
16\pi^2 \frac{d Y_d}{d\log\m}&= \left[ -\frac{3}{2} Y_u Y_u^\dagger  + \frac{3}{2} Y_d Y_d^\dagger 
+ 3 {\tr} \left(Y_u Y_u^\dagger + Y_d Y_d^\dagger \right)-\frac{5}{12}g'^2 - \frac{9}{4}g_2^2 - 8 g_s^2 \right] Y_d. \label{eq:RGE_yd}
\end{align}
Here, we neglect the lepton Yukawa interactions. 
The RG equations for $g'$, $Y_u$ and $Y_d$ at one-loop level are same as the SM ones \cite{Machacek:1981ic}.
RG equations for the Wilson coefficients are given by,\footnote{
One may worry about mixing with other operators, such as,
$d_{abc} ({\tilde B}^\dagger \bar\s^\m \tilde g^a)({\tilde g}^{b\dagger} \bar\s_\m \tilde g^c)$ and 
$d_{abc} ({\tilde B} \tilde g^a)({\tilde g}^b \tilde g^c)$.  Here, $d_{abc} = {\rm tr}[ T^a \{T^b,T^c\}]$.
However, $Q_{Q,ij}^{\tilde B}$, $Q_{u,ij}^{\tilde B}$, $Q_{d,ij}^{\tilde B}$ and $Q_7^{\tilde B}$ do not mix with them at one-loop level.
}
\begin{align}
16\pi^2 \frac{d C^{\tilde B}_Q}{d\log\m} &= \frac{1}{2} \left\{ Y_u Y_u^\dagger + Y_d Y_d^\dagger, C^{\tilde B}_Q \right\} - Y_u C^{\tilde B}_u Y_u^\dagger - Y_d C^{\tilde B}_d Y_d^\dagger- 3 g_s^2 N_C C^{\tilde B}_Q + \frac{2}{3} g_s^2 \hat {\bf S} , \label{eq:RGE_bino_q}\\
16\pi^2 \frac{d C^{\tilde B}_u}{d\log\m} &= \left\{ Y_u^\dagger Y_u, C^{\tilde B}_u \right\} - 2 Y_u^\dagger C^{\tilde B}_Q Y_u - 3 g_s^2 N_C C^{\tilde B}_u + \frac{2}{3} g_s^2 
\hat {\bf S},  \label{eq:RGE_bino_u}\\
16\pi^2 \frac{d C^{\tilde B}_d}{d\log\m} &= \left\{ Y_d^\dagger Y_d, C^{\tilde B}_d \right\} 
- 2 Y_d^\dagger C^{\tilde B}_Q Y_d - 3 g_s^2 N_C C^{\tilde B}_d + \frac{2}{3} g_s^2 \hat{\bf S} , \label{eq:RGE_bino_d}\\
16\pi^2 \frac{d C^{\tilde B}_7}{d\log\m} &= \left( \frac{2}{3}N_F - 6N_C \right) g_s^2 C^{\tilde B}_7, \label{eq:RGE_bino_7}\\
16\pi^2 \frac{d C^{\tilde W}}{d\log\m} &= \frac{1}{2} \left\{ Y_u Y_u^\dagger + Y_d Y_d^\dagger, C^{\tilde W} \right\} - 3 g_s^2 N_C C^{\tilde W} - 6 g^2 C^{\tilde W}. \label{eq:RGE_wino}
\end{align}
Here, $\hat{\bf S} = ( 2{\rm tr} C^{\tilde B}_Q + {\rm tr} C^{\tilde B}_u + {\rm tr} C^{\tilde B}_d) {\bf 1}  $, $\{ A,B \} = AB + BA$, ${\bf 1} = {\rm diag}(1,1,1)$, $N_F = 6$ and $N_C = 3$. 
Again, $C^{\tilde B}_Q$, $C^{\tilde B}_u$, $C^{\tilde B}_d$ and $C^{\tilde W}$ are $3\times 3$ matrices in flavor space.
We show one-loop diagrams which contribute to the above RG equations in Fig. \ref{fig:RGdiagram}.
We have checked Eqs. (\ref{eq:RGE_bino_q}--\ref{eq:RGE_wino}) are consistent with Ref. \cite{Gambino:2005eh} in the flavor-symmetric case.~\footnote{
If the Higgsino is lighter than the gluino, we have to consider effective operators which include Higgsinos.
For RG equation of their Wilson coefficients, see the Appendix \ref{sec:higgsinoRG}.}
In the following of this section, we will discuss the flavor structures of the Wilson coefficients in more detail.

\subsection{Resummed Wilson Coefficient}
\begin{figure}[t]
\centering
\includegraphics[width=0.5\hsize]{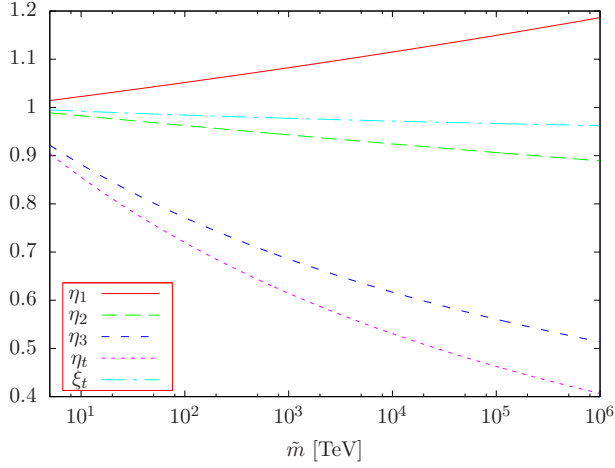}
\caption{
Variables $\eta_1$, $\eta_2$, $\eta_3$, $\eta_t$ and $\xi_t$  in Eqs. (\ref{eq:functions_}--\ref{eq:xit}) as 
functions of $\tilde m$, fixing $\mu=|m_{\tilde g}|$.
We take $m_{\tilde g}=1.5$ TeV and $m_{\tilde W} = 200$ GeV.
}\label{fig:coupling}
\end{figure}

Here, we show an analytic solution for an approximated situation.
We neglect the Yukawa couplings except for the top Yukawa, namely, we take the following Yukawa matrices,
\begin{eqnarray}
Y_u ~=~ \left(\begin{array}{c c c}
0&0&0\\
0&0&0\\
0&0&y_t\\
\end{array}\right), ~~~~~~~~~~
Y_d ~=~ \left(\begin{array}{c c c}
0&0&0\\
0&0&0\\
0&0&0\\
\end{array}\right). \label{eq:yukawa_simple}
\end{eqnarray}
To write down the solution to the RG equations, we define the following variables:\footnote{In the limit of $g'=g=0$, we can get analytic form of $\eta_t$:
\begin{eqnarray}
\eta_t &=& \eta_3^{8/5} - \frac{3}{2} \frac{\a_t(\tilde m)}{\a_s(\tilde m)} \left(\eta_3^{8/5}-\eta_3 \right)=
 \left(\eta_3^{-8/5} - \frac{3}{2} \frac{\a_t(\m)}{\a_s(\m)} \left(\eta_3^{-8/5}-\eta_3^{-1} \right)\right)^{-1}.
\end{eqnarray}
}
\begin{eqnarray}
\eta_1 &\equiv& \frac{\a'(\tilde m)}{\a'(\m)} ~=~ 1 - \frac{41 \a'(\tilde m)}{12\pi}\log\frac{\m}{\tilde m} = 
 \left(1 + \frac{41 \a'(\m)}{12\pi}\log\frac{\m}{\tilde m}\right)^{-1}, \label{eq:functions_}\\
\eta_2 &\equiv& \frac{\a(\tilde m)}{\a(\m)} ~=~ 1 + \frac{11 \a(\tilde m)}{12\pi}\log\frac{\m}{\tilde m}=
 \left(1 - \frac{11 \a(\m)}{12\pi}\log\frac{\m}{\tilde m}\right)^{-1},
 \\
\eta_3 &\equiv& \frac{\a_s(\tilde m)}{\a_s(\m)} ~=~ 1 + \frac{5 \a_s(\tilde m)}{2\pi}\log\frac{\m}{\tilde m}
=\left(1 - \frac{5 \a_s(\m)}{2\pi}\log\frac{\m}{\tilde m}\right)^{-1}
, \\
\eta_t &\equiv& \frac{\a_t(\tilde m)}{\a_t(\m)}, \\
\xi_t &\equiv& \exp\left( -\int^{\tilde m}_\m \frac{y_t^2(\m')}{16\pi^2} \frac{d\m'}{\m'} \right) ~=~ \eta_t^{-1/9} \eta_1^{-17/738} \eta_2^{3/22} \eta_3^{8/45} \label{eq:xit}, \\
y &\equiv& \eta_3^{4/5} - 1, \\
z &\equiv& \frac{1}{3}\left(\xi_t^{3} - 1 \right), \\
z' &\equiv& \frac{1}{3}\left(\xi_t^{3/2} - 1 \right), \\
\bar C^{\tilde B}(\tilde m) &\equiv& \frac{1}{12} \left( 2{\rm tr} C^{\tilde B}_Q(\tilde m) + {\rm tr} C^{\tilde B}_u(\tilde m) + {\rm tr} C^{\tilde B}_d(\tilde m) \right). \label{eq:functions__}
\end{eqnarray}
We show $\eta_1$, $\eta_2$, $\eta_3$, $\eta_t$ and $\xi_t$ as functions of $\tilde m$ with $\mu=|m_{\tilde g}|$ in Fig. \ref{fig:coupling}.
Here we set $m_{\tilde g}=1.5$ TeV, $m_{\tilde W} = 200$ GeV and $m_{\tilde B} =400$ GeV.\footnote{
Very recently, a search of disappearing charged tracks by the ATLAS collaboration gives 
severe constraint on the mass of the Wino LSP.
If the Higgsino mass is large enough and we adopt two-loop level mass splitting between $\tilde W^{\pm}$ and $\tilde W^0$ \cite{Ibe:2012sx},
we obtain $m_{\tilde W} > 270~{\rm GeV}$ \cite{winodirect}. 
However, our analysis is almost independent on the Wino mass, then, we take this value as a reference point.
}
By using these variables given in Eqs. (\ref{eq:functions_}--\ref{eq:functions__}),
the solution to the RG equations in Eqs. (\ref{eq:RGE_bino_q}--\ref{eq:RGE_wino}) is written as,
\begin{eqnarray}
C^{\tilde W}_{ij}(\m) &=& \eta_3^{-9/10} \eta_2^{-18/11}~ C^{\tilde W}_{ij}(\tilde m), \label{eq:CWij}\\
C^{\tilde W}_{3i}(\m) &=& \eta_3^{-9/10} \eta_2^{-18/11}~ \xi_t^{1/2} C^{\tilde W}_{3i}(\tilde m), \label{eq:CW3i}\\
C^{\tilde W}_{33}(\m) &=& \eta_3^{-9/10} \eta_2^{-18/11}~ \xi_t C^{\tilde W}_{33}(\tilde m), \label{eq:CW33} \\[5mm]
%\end{eqnarray}
%
%\begin{eqnarray}
C^{\tilde B}_{Q,ii}(\m) &=& \eta_3^{-9/10}~ \left[ C^{\tilde B}_{Q,ii}(\tilde m) + y \bar C^{\tilde B}(\tilde m) \right], \\
C^{\tilde B}_{Q,12}(\m) &=& \eta_3^{-9/10}~ C^{\tilde B}_{Q,12}(\tilde m), \\
C^{\tilde B}_{Q,33}(\m) &=& \eta_3^{-9/10}~ \left[ (1+z)C^{\tilde B}_{Q,33}(\tilde m) -z C^{\tilde B}_{u,33}(\tilde m) + y \bar C^{\tilde B}(\tilde m) \right], \\
C^{\tilde B}_{Q,3i}(\m) &=& \eta_3^{-9/10}~ \left[ (1+z')C^{\tilde B}_{Q,3i}(\tilde m) -z' C^{\tilde B}_{u,3i}(\tilde m) \right],\\[5mm]
%\end{eqnarray}
%
%\begin{eqnarray}
C^{\tilde B}_{u,ii}(\m) &=& \eta_3^{-9/10}~ \left[ C^{\tilde B}_{u,ii}(\tilde m) + y \bar C^{\tilde B}(\tilde m) \right], \\
C^{\tilde B}_{u,12}(\m) &=& \eta_3^{-9/10}~ C^{\tilde B}_{u,12}(\tilde m), \\
C^{\tilde B}_{u,33}(\m) &=& \eta_3^{-9/10}~ \left[ (1+2z)C^{\tilde B}_{u,33}(\tilde m) -2z C^{\tilde B}_{Q,33}(\tilde m) + y \bar C^{\tilde B}(\tilde m) \right], \\
C^{\tilde B}_{u,3i}(\m) &=& \eta_3^{-9/10}~ \left[ (1+2z')C^{\tilde B}_{u,3i}(\tilde m) -2z' C^{\tilde B}_{Q,3i}(\tilde m) \right], \\[5mm]
%\end{eqnarray}
%
%\begin{eqnarray}
C^{\tilde B}_{d,ii}(\m) &=& \eta_3^{-9/10}~ \left[ C^{\tilde B}_{d,ii}(\tilde m) + y \bar C^{\tilde B}(\tilde m) \right], \\
C^{\tilde B}_{d,12}(\m) &=& \eta_3^{-9/10}~ C^{\tilde B}_{d,12}(\tilde m), \\
C^{\tilde B}_{d,33}(\m) &=& \eta_3^{-9/10}~ \left[ C^{\tilde B}_{d,33}(\tilde m) + y \bar C^{\tilde B}(\tilde m) \right], \\
C^{\tilde B}_{d,3i}(\m) &=& \eta_3^{-9/10}~ C^{\tilde B}_{d,3i}(\tilde m), \\[5mm]
%\end{eqnarray}
%
%\begin{eqnarray}
C^{\tilde B}_7(\m) &=& \eta_3^{-7/5}~ C^{\tilde B}_7(\tilde m).
\end{eqnarray}
Here, $i$ runs a light flavor index $1,2$. 
Note that we have dropped off-diagonal elements of Yukawa matrices.
Then, off-diagonal elements of $C$'s are valid only if the contribution of off-diagonal elements of Yukawa is much smaller than them.
For example, $C^{\tilde W}_{12}(\m)$ in Eq. (\ref{eq:CWij}) is valid
only if $C^{\tilde W}_{12}(\tilde m)$ is much larger than $Y_{u,12} Y_{u,22}C^{\tilde W}_{22}(\tilde m)$ and $Y_{u,13} Y_{u,23}C^{\tilde W}_{33}(\tilde m)$.

\subsection{Numerical Analysis of Gluino decay}
So far, we discussed leading logarithm corrections for the Wilson coefficients relevant to the gluino decay.
To extract information on the squark sector from the gluino decay, it is important to take into account such corrections.
In this subsection, we discuss the effects of leading logarithm corrections on the lifetime and decay pattern of the gluino numerically.
In the present situation, gluino two-body decay ($\tilde g\to g \tilde B$) can be neglected because of one-loop suppression.\footnote{
If the Higgsino is lighter than the gluino, the decay width of this two body decay mode is enhanced because of large logarithmic corrections \cite{Toharia:2005gm, Gambino:2005eh, Sato:2012xf}. Then, this mode can be significant.
}
In the limit that $|m_{\tilde g}|-|m_{\tilde W}|,~|m_{\tilde g}|-|m_{\tilde B}| \gg m_t$ and $V_{\rm CKM} \simeq {\bf 1}$,
we can approximately get the following analytical formulae \cite{Gambino:2005eh, Barbieri:1987ed}:
\begin{align}
\G(\tilde g \to \tilde B u_{Li} \bar u_{L j}) =
\G(\tilde g \to \tilde B d_{Li} \bar d_{L j}) &= \frac{|C_{Q,ij}^{\tilde B}|^2 |m_{\tilde g}|^5}{1536\pi^3 {\tilde m}^4} f\left( m_{\tilde B} / m_{\tilde g} \right), \\
\G(\tilde g \to \tilde B u_{R i} \bar u_{R j}) &= \frac{|C_{u,ij}^{\tilde B}|^2 |m_{\tilde g}|^5}{1536\pi^3 {\tilde m}^4} f\left( m_{\tilde B} / m_{\tilde g} \right), \\
\G(\tilde g \to \tilde B d_{R i} \bar d_{R j}) &= \frac{|C_{d,ij}^{\tilde B}|^2 |m_{\tilde g}|^5}{1536\pi^3 {\tilde m}^4} f\left( m_{\tilde B} / m_{\tilde g} \right), \\
\G(\tilde g \to \tilde W^0 u_{Li} \bar u_{L j}) =
\G(\tilde g \to \tilde W^0 d_{Li} \bar d_{L j}) &= \frac{|C_{ij}^{\tilde W}|^2 |m_{\tilde g}|^5}{1536\pi^3 {\tilde m}^4} f\left( m_{\tilde W} / m_{\tilde g} \right), \\
\G(\tilde g \to \tilde W^+ d_{Li} \bar u_{L j}) =
\G(\tilde g \to \tilde W^- u_{Li} \bar d_{L j}) &= \frac{2|C^{\tilde W}_{ij}|^2 |m_{\tilde g}|^5}{1536\pi^3 {\tilde m}^4} f\left( m_{\tilde W} / m_{\tilde g} \right).
\end{align}
Here, $f(x) = 1-8|x|^2-12|x|^4\log |x|^2 + 8|x|^6 - |x|^8 + 2(1+9|x|^2+6|x|^4\log |x|^2 - 9|x|^4 + 6|x|^4\log |x|^2-|x|^6) {\rm Re}(x) $.
As for operators whose dimension is higher than six, we use tree-level amplitudes in our numerical calculation.
In the following of this section, we denote decay widths calculated by $C(\m=\tilde m)$ and  $C(\m=|m_{\tilde g}|)$ 
as ``tree'' and  ``resum'', respectively.

\subsubsection*{Total decay width}
\begin{figure}
\begin{center}
\includegraphics[width=0.6\hsize]{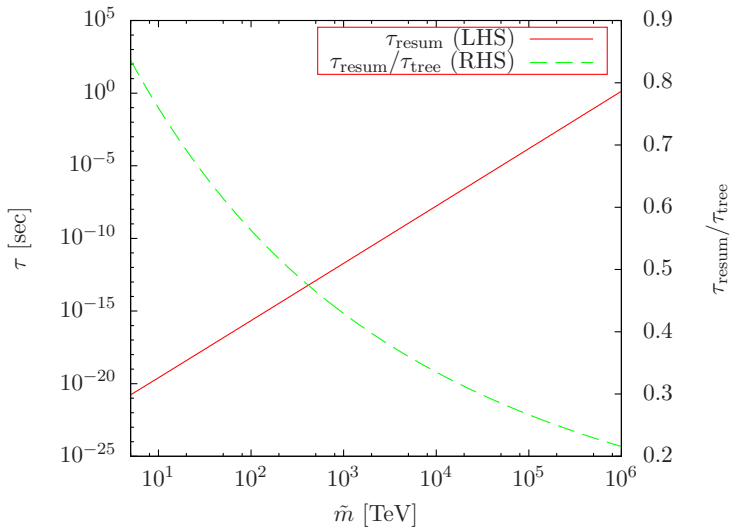}
\caption{
The red and solid line shows the lifetime of the gluino with resummation.
The green and dashed line shows the ratio of the resummed lifetime to tree-level one.
We take the gaugino masses as $m_{\tilde W} = 200~\GEV$, $m_{\tilde B} = 400~\GEV$ and $m_{\tilde g} = -1.5~\TEV$.
For the squark mass, we take universal squark masses at the squark mass scale.
}\label{fig:lifetime}
\end{center}
\end{figure}

It is pointed out that the lifetime of the gluino is significantly affected by the resummation of leading logarithmic corrections \cite{Gambino:2005eh}.
We show the lifetime of the gluino in Fig. \ref{fig:lifetime}.
Here we take $m_{\tilde Q_L}^2 = m_{\tilde u_R}^2 = m_{\tilde d_R}^2 = \tilde m^2 \times {\bold 1} $.
We can see that the quantum corrections make the decay rate double for  $\tilde m=10^3$ TeV.

\subsubsection*{Branching fraction of $\tilde g \to \tilde W q\bar q$ and $\tilde g \to \tilde B q\bar q$}
In the case of the heavy Higgsino, 
the low-mass neutralinos and charginos are almost pure Wino or Bino states.
Therefore discriminating the gluino decay mode into a Wino and Bino can be experimentally viable.
This branching ratio reveals the relation between left-handed squarks and right-handed ones.
Thus, precise estimation of the branching ratio is significant.

At tree level, the Wilson coefficients of the operators involving $\tilde W$ are proportional to the gauge coupling $g$,
and those involving $\tilde B$ are proportional to $g'$.
They are proportional to $\a'$ and $\a$ at tree level, 
however, leading log resummation alters ratio between them.

\begin{figure}
\centering
\includegraphics[width=0.6\hsize]{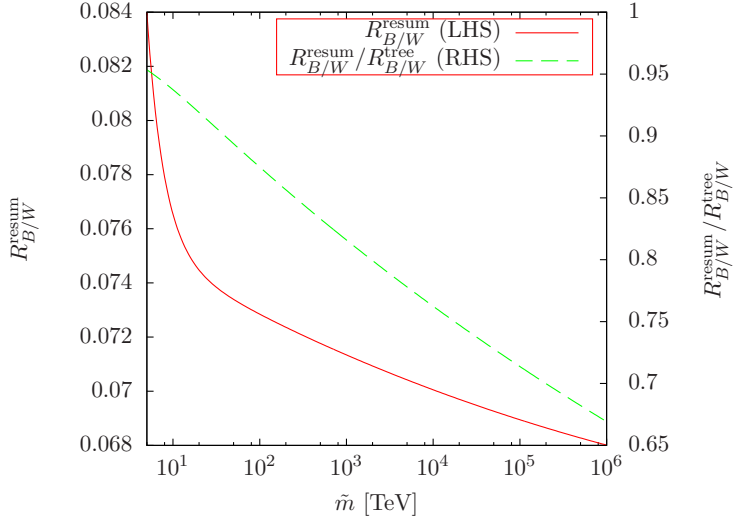}
\caption{
$R_{\tilde B/ \tilde W}|_{\rm resum}$ (red and solid line) and the ratio of the resummed one to tree-level one (green and dashed line).
We take the gaugino masses as $m_{\tilde W} = 200~\GEV$, $m_{\tilde B} = 400~\GEV$ and $m_{\tilde g} = -1.5~\TEV$.
For the squark mass, we take universal squark mass at for $\tilde m=10^3$ TeV.quark mass scale.
  }\label{fig:binowino}
\end{figure}

In Fig. \ref{fig:binowino}, we plot $R_{\tilde B / \tilde W}$ as a function of squark masses, which is defined as,
\begin{eqnarray}
R_{\tilde B / \tilde W} &\equiv&
\frac{ \sum_q \G(\tilde g \to \tilde B q\bar q) }{ \sum_q \G(\tilde g \to \tilde W q\bar q)}.
\end{eqnarray}
Here we take $m_{\tilde Q_L}^2 = m_{\tilde u_R}^2 = m_{\tilde d_R}^2 = \tilde m^2 \times {\bold 1} $.
We can see that resummation of leading logarithm corrections alters branching fractions by about 20 \% for $\tilde m=10^3$ TeV.

\subsubsection*{Flavor structure of $\tilde g \to \tilde W q\bar q$}
The flavor structure of $\tilde g \to \tilde W q\bar q$ is determined by $(m^2_{\tilde Q_L})^{-1}$ at tree level,
and it suffers from leading logarithm corrections because of the anomalous dimension of the quarks.
An important source of the corrections is top Yukawa $y_t$.
It changes $\tilde g \to \tilde W^0 t\bar t$ relative ratio to $\tilde g \to \tilde W q \bar q$.
\begin{eqnarray}
R^{\tilde W}_{tt / qq} &\equiv& \frac{ \G(\tilde g \to \tilde W^0 t\bar t )}{ \G(\tilde g \to \tilde W q_i\bar q_j)},
\end{eqnarray}
where $i,j=1,2$.
The ratio of resummed result to tree-level one is $(R^{\tilde W}_{tt/qq}|_{\rm resum}) / (R^{\tilde W}_{tt/qq}|_{\rm tree}) = \xi_t^2$.
As seen in Fig. \ref{fig:coupling},  $R^{\tilde W}_{tt / qq}$ is 
decreased by the effect of resummation of leading logarithm correction by 3--4 \%, compared to the tree-level result.

\subsubsection*{Flavor structure of $\tilde g \to \tilde B q\bar q$}
Here, we discuss radiative correction for the following quantities:
\begin{eqnarray}
R^{\tilde B}_{tc/tt} ~=~ \frac{ \G(\tilde g \to \tilde B t\bar c/ \tilde B \bar t c )}{ \G(\tilde g \to \tilde B t\bar t)},
\quad\quad\quad
R^{\tilde B}_{tc/cc} ~=~ \frac{ \G(\tilde g \to \tilde B t\bar c/ \tilde B \bar t c )}{ \G(\tilde g \to \tilde B c\bar c)}.
\end{eqnarray}
Compared to the Wino case, radiative corrections for the decay into the Bino more significantly affect the flavor structure of the gluino decay.
This is because the contribution of the top Yukawa coupling is larger.
In addition, the penguin-like diagram contribution in Fig.~\ref{fig:RGdiagram} affects only the flavor diagonal part.
Thus, the RG evolution of the Wilson coefficients has different feature between the flavor conserving and violating parts.
We show numerical results in the following two sample cases.

\begin{itemize}
\begin{figure}
\centering
\includegraphics[width=\hsize]{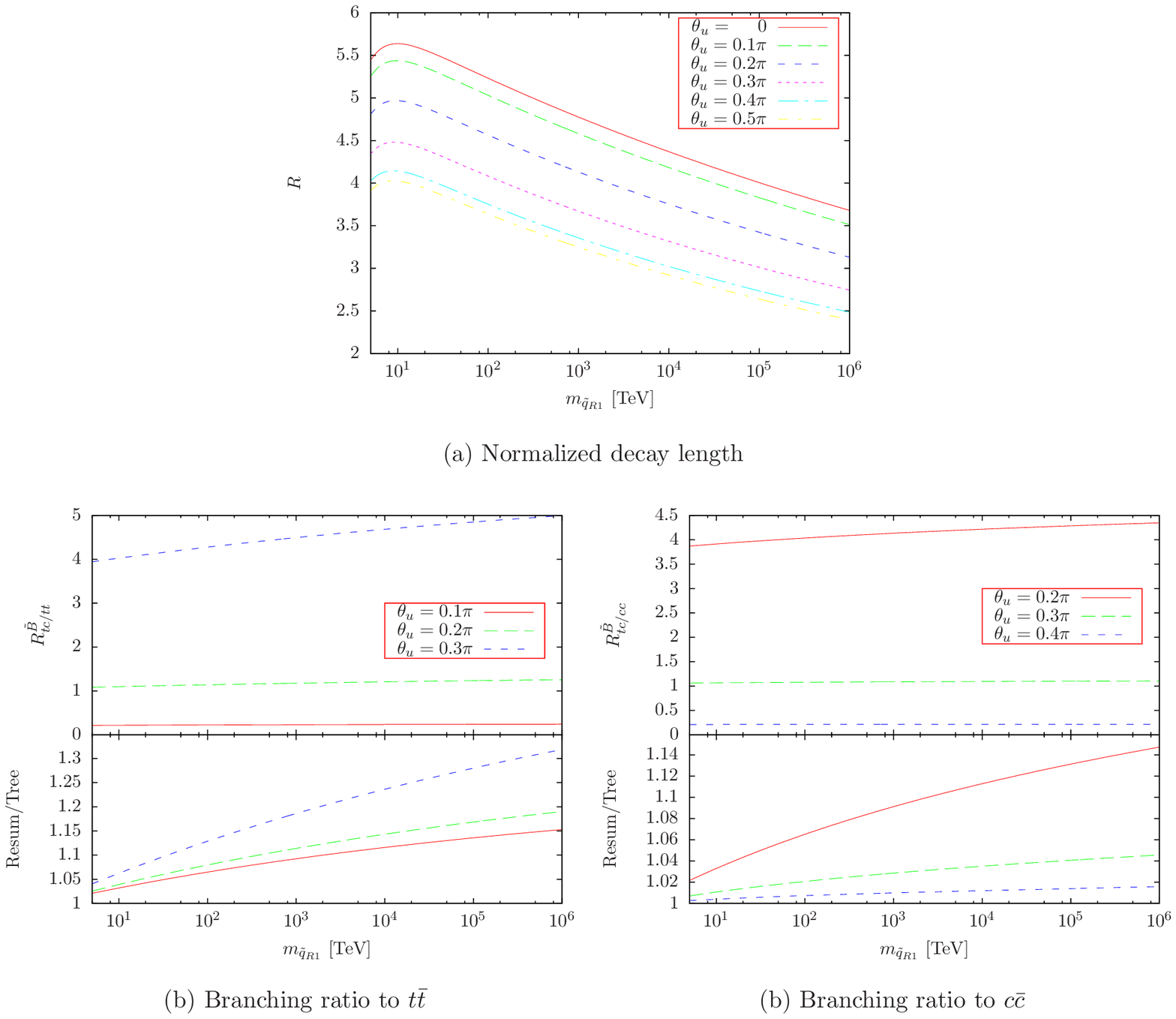}
\caption{
Normalized decay length which is defined as $R = (c\tau_{\tilde g} / 1~{\rm cm}) (m_0 / 1000~\TEV)^4$ (a),
branching ratio $R^{\tilde B}_{tc/tt}$ (b) and $R^{\tilde B}_{tc/cc}$ (c) in a case of light $\tilde u_R$.
We take the gaugino masses as $m_{\tilde W} = 200~\GEV$, $m_{\tilde B} = 400~\GEV$ and $m_{\tilde g} = -1.5~\TEV$.
}\label{fig:udominate}
\end{figure}

\item light $\tilde u_R$ \\
In this case, we assume that the lightest squark $\tilde q_{R1}$ is a mixture of $\tilde t_R$ and $\tilde c_R$, and other squarks are much heavier than them.
The mixing angle $\theta_u$ is defined as $\tilde q_{R1} = \cos\theta_u \tilde t_R + \sin\theta_u \tilde c_R$.
Fig. \ref{fig:udominate} shows the numerical result of the gluino decay length, $R^{\tilde B}_{tc/tt}$ and $R^{\tilde B}_{tc/cc}$.
Here we define a normalization variable $R$ for the gluino decay length as
\begin{align}
c\tau_{\tilde g} = R  \times 1~{\rm cm} \left(\frac{m_{\tilde q_{R1}}}{ 1000~\TEV} \right)^4.
\end{align}
We can see $R^{\tilde B}_{tc/tt}|_{\rm resum}$ ($R^{\tilde B}_{tc/cc}|_{\rm resum}$) is enhanced compared to $R^{\tilde B}_{tc/tt}|_{\rm tree}$ ($R^{\tilde B}_{tc/cc}|_{\rm tree}$) results by 10--20 \%.

\item mSUGRA-like case\\
In this case, at the ``GUT'' scale $\m=2\times 10^{16}$ GeV, we take the
diagonal elements of the squark mass-squared as $m_{\tilde Q_L,ii}^2 = m_{\tilde u_R,ii}^2 = m_{\tilde d_R,ii}^2 = m_0^2$ ($i=1,2,3$),
the off-diagonal elements as $m_{\tilde u_R,23}^2 = m_{\tilde u_R,32}^2 = m_{23}^2$ and the other components to be zero.
For other parameters, we take $m_{\tilde W} = 200~\GEV,~ m_{\tilde B} = 400~\GEV,~m_{\tilde g} = -1.5~\TEV$ and $\tan\b = 1, 3$.
By using RG equations for the squark soft parameters, we obtain physical squark masses and mixings.
In calculation of soft mass RG evolution, we neglect terms other than top Yukawa parts.
We also assume $b$, $\mu$ and $m_{H_d}^2$ are optimized so that the correct electroweak breaking occurs.

For estimation of the gluino decay width, we take the matching scale $\tilde m$ as the lightest squark mass.
In this case, the RG effect from the top Yukawa coupling decreases $m_{\tilde t_R}$ and the decay into $t\bar t \tilde B$ is enhanced.
Fig. \ref{fig:msugralike} shows the numerical result of $R^{\tilde B}_{tc/tt}$.
The shaded region shows the region where the lightest squark is lighter than the gluino or a tachyon.
We can see $R^{\tilde B}_{tc/tt}|_{\rm resum}$ ($R^{\tilde B}_{tc/cc}|_{\rm resum}$) is 
enhanced compared to $R^{\tilde B}_{tc/tt}|_{\rm tree}$ ($R^{\tilde B}_{tc/cc}|_{\rm tree}$) results by about 10 \% for $m_0 ={\cal O}(10^3)$ TeV.
\begin{figure}[htbp]
\centering
\includegraphics[width=0.9\hsize]{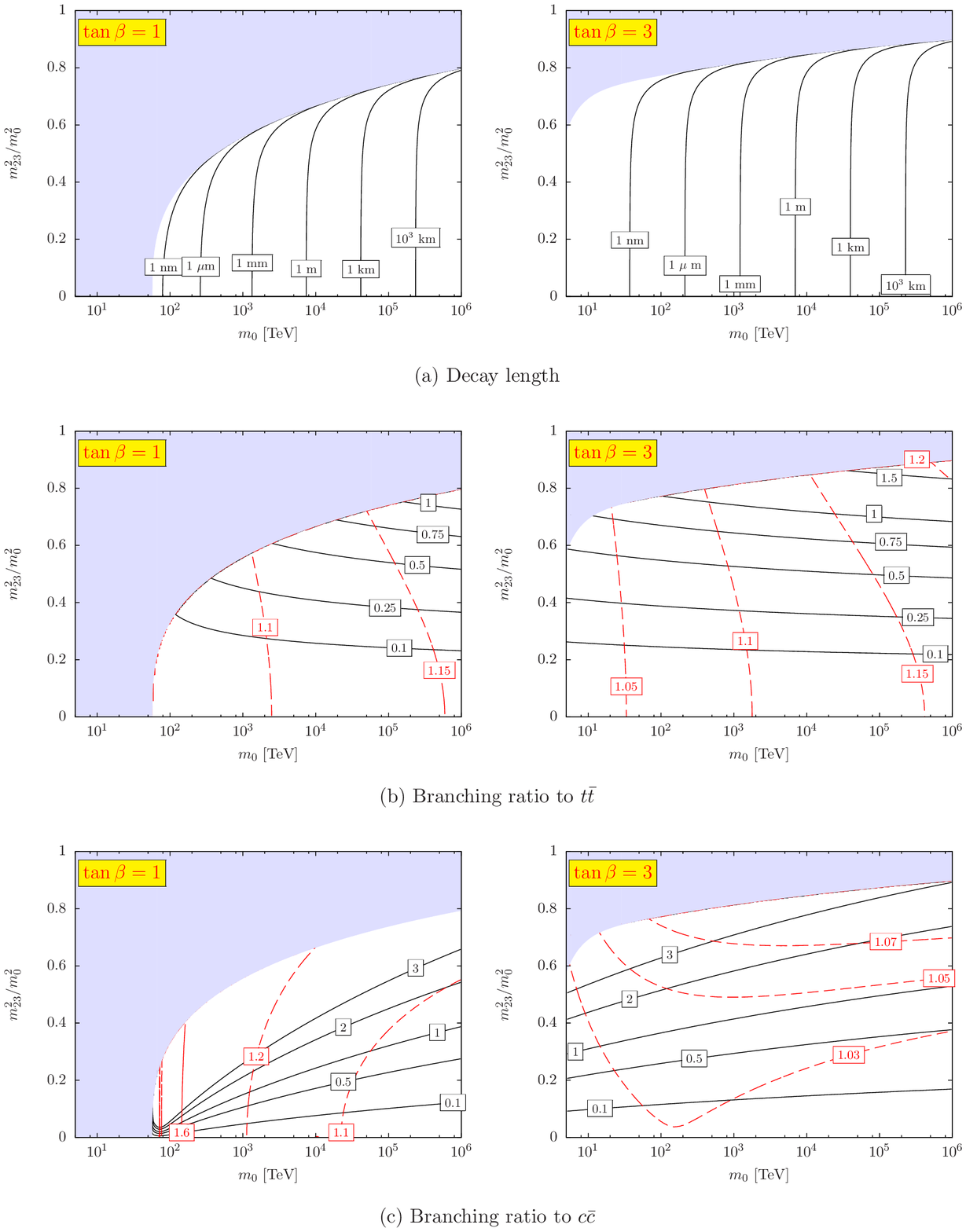}
\caption{
Decay length $c\tau_{\tilde g}$ (a),
Branching fraction $R^{\tilde B}_{tc/tt}$ (b) and $R^{\tilde B}_{tc/cc}$ (c) in mSUGRA-like case.
In shaded region of the above figures, the lightest squark becomes lighter than the gluino or a tachyon.
In figures (b) and (c), black lines show branching ratio $R^{\tilde B}_{tc/tt}$ and $R^{\tilde B}_{tc/cc}$, respectively, and red-dashed lines shows each $R_{\rm resum} / R_{\rm tree}$.
}\label{fig:msugralike}
\end{figure}

\end{itemize}
%

%%%%%%%%%%%%%%%%%%%%%%%%%%%%%%%%%%%%%%%%%%%%%%%%%%%%%%%%
% Flavor constraints
%%%%%%%%%%%%%%%%%%%%%%%%%%%%%%%%%%%%%%%%%%%%%%%%%%%%%%%%

\section{Low-energy Flavor Constraints and Gluino Decay}
As we have discussed above, the gluino decay directly reflects the structure of the squark sector.
For example, the flavor violation of the squark sector leads the flavor violating gluino decay, such as $\tilde g \to t c \tilde \chi$.
However, in general, the gluino-squark interactions which control the gluino decay, also contribute to 
low-energy rare processes such as meson mixings and electric dipole moments (EDM's).
Therefore, observations of the low-energy physics give constraints on the gluino decay process.
In this section, we study the correlation between the low-energy flavor/CP observation and the gluino decay.

\subsection{Flavor/CP Constraints}
In general, SSMs can give visible contributions to low-energy flavor/CP violating processes, even if the SUSY breaking scale is
high, say, higher than ${\cal O}(10)$ TeV.
Here we discuss some low-energy observables, which can give strong constraints on the high-scale SUSY models.
For reviews, see, e.g., Refs. \cite{Gabbiani:1996hi,Altmannshofer:2009ne}.

Here we adopt the super-CKM basis (see Appendix \ref{sec:notation} for the correspondence to the weak-basis).
We consider the following squark mass-squared matrices in this basis:
\begin{equation}
{m}^2_{a} \!=\! {m}_0^2 
\left( 
\begin{matrix}
1+\Delta_1^a & \delta_{12}^{a} & \delta_{13}^{a}\\
\delta_{12}^{a*} & 1+\Delta_2^a & \delta_{23}^{a}\\
\delta_{13}^{a*}  & \delta_{23}^{*}  & 1+\Delta_3^a\\
\end{matrix}
\right),
\end{equation}
where $a = \tilde{Q}_L (\tilde{d}_L), \tilde{u}_R, \tilde{d}_R$. 
The mass-squared matrix for $\tilde{u}_L$ is obtained by $\hat{m}^2_{\tilde{u}_L}=V_{\rm CKM}\ {m}^2_{\tilde{Q}_L} V_{\rm CKM}^\dagger$. 
For the assumption of some charge of the SUSY breaking field, 
we expect the sizes of $A$-terms are order of the gaugino masses and neglect them.
We discuss the current constraints on the flavor violating mass terms.
\subsection*{$\Delta F=2$ process}

\begin{figure}[htbp]
\begin{center}
\includegraphics[width=120mm]{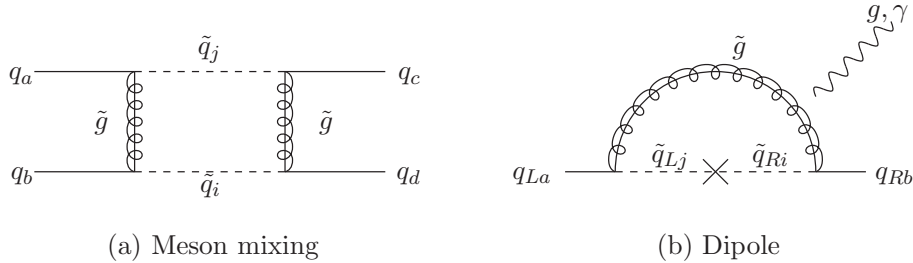}
\caption{Examples of diagrams relevant to low-energy constraints.
}
\label{fig:loop}
\end{center}
\end{figure}
One of the strongest constraints on  $\delta$'s comes from $\Delta F=2$ meson mixings.
Squarks contribute to the low-energy $\Delta F = 2$ operators via loop processes (Fig. \ref{fig:loop}-(a)).
We consider only gluino mediated box diagrams, which give leading contributions.
The $K^0-\bar K^0 $ system is well measured and gives the strongest constraint on the first and second generation flavor violation.
This can constrain $\delta_{12}^{\tilde Q_L}$ and $\delta_{12}^{\tilde d_R}$, since they induce direct $s -d$ mixings. 
Especially, simultaneous existences of $\delta_{12}^{\tilde Q_L}$, $\delta_{12}^{\tilde d_R}$ and CP-phase are drastically constrained.
In addition, it can also constrain products of $\delta_{13}$ and $\delta_{23}$.
This is because a squark transition $(1 \to 3 \to 2)$ induces large $s -d$ transition.

\subsection*{$\Delta F=1$ process}
The operators with $\Delta F=1$ induce $\Delta F=1$ heavy quark decays, such as $b\to s \gamma$.
These operators are chirality-flipping, and thus the amplitudes of these processes depend on
the size of $\mu$ term.
If the absolute value of $\mu$ is as much as the squark scale, the observation from $b\to s \gamma$ provides a strong constraint
on the entry of 3-2 element of the squark mass-squared matrices.
In the case of ${\cal O}(1)$ of $\delta^{23}$, $m_0 < {\cal O}(10)$ TeV is excluded.

\subsection*{$\Delta F=0$ process}
Electric and chromo-electric dipole moments (EDM's and CEDM's) of quarks give constraints on the sfermion sector.
Although the EDM and CEDM are not flavor violating processes, 
it can severely constrain the squark flavor violating structure, if both left and right-handed squarks 
sector have flavor/CP violations and the size of $\mu$ is large.
These operators are chirality-flipping and the coefficients are suppressed by the SM fermion masses.
When the squark mass-squared matrix is flavor diagonal, these amplitudes are suppressed by the light quark mass ($m_u$, $m_d$ and $m_s$).
However once one allows large flavor violation in the squark sector,  the suppression is replaced with heavy quarks ($m_t$ and $m_b$) and
the size of EDM enhanced by $m_{t,b}/m_{u,d,s}$.
With ${\cal O}(1)$ flavor violation, even PeV-scale SUSY can provide significant EDM signatures as emphasized in Ref. \cite{McKeen:2013dma}.

\subsection*{Current Constraints}
\begin{figure}[htbp]
\begin{center}
\subfigure[CP + R]{\includegraphics[width=75mm]{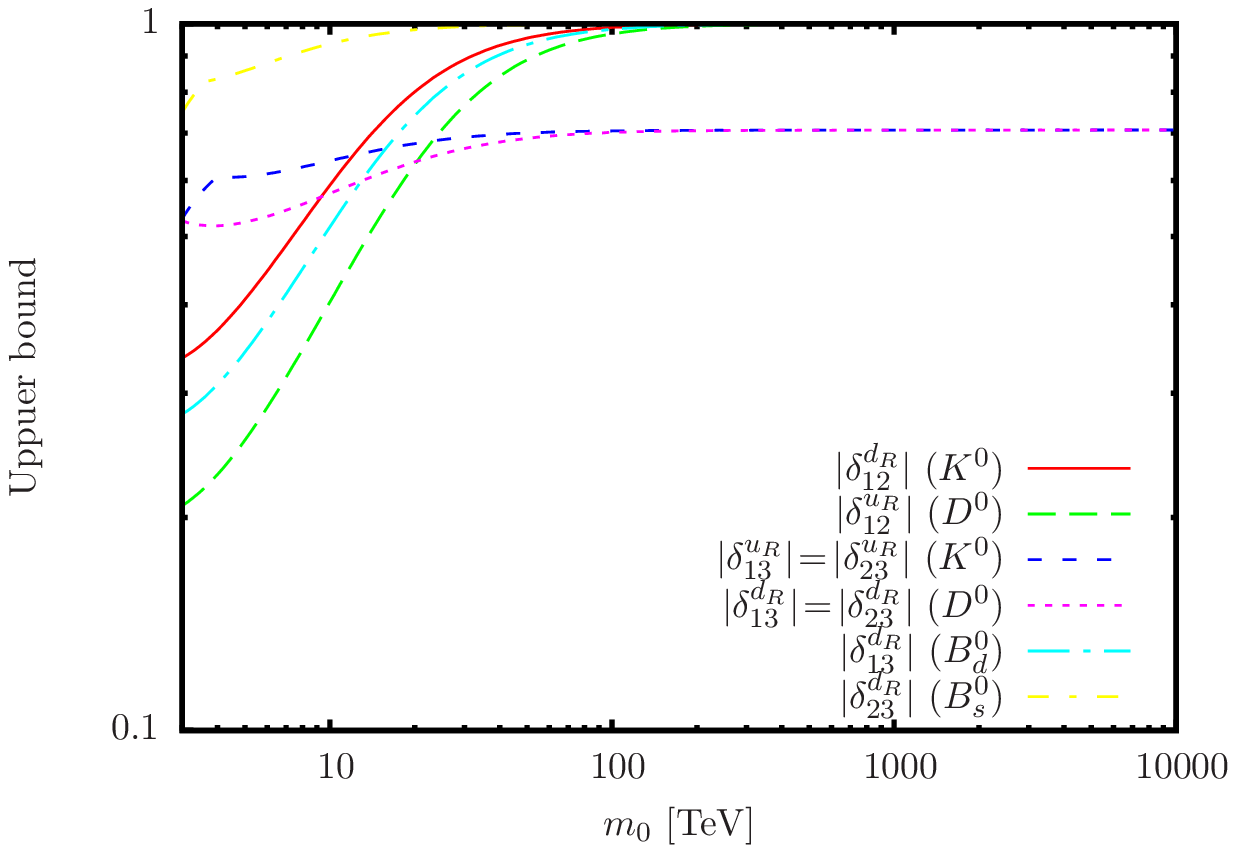}}
\subfigure[CP + LR]{\includegraphics[width=75mm]{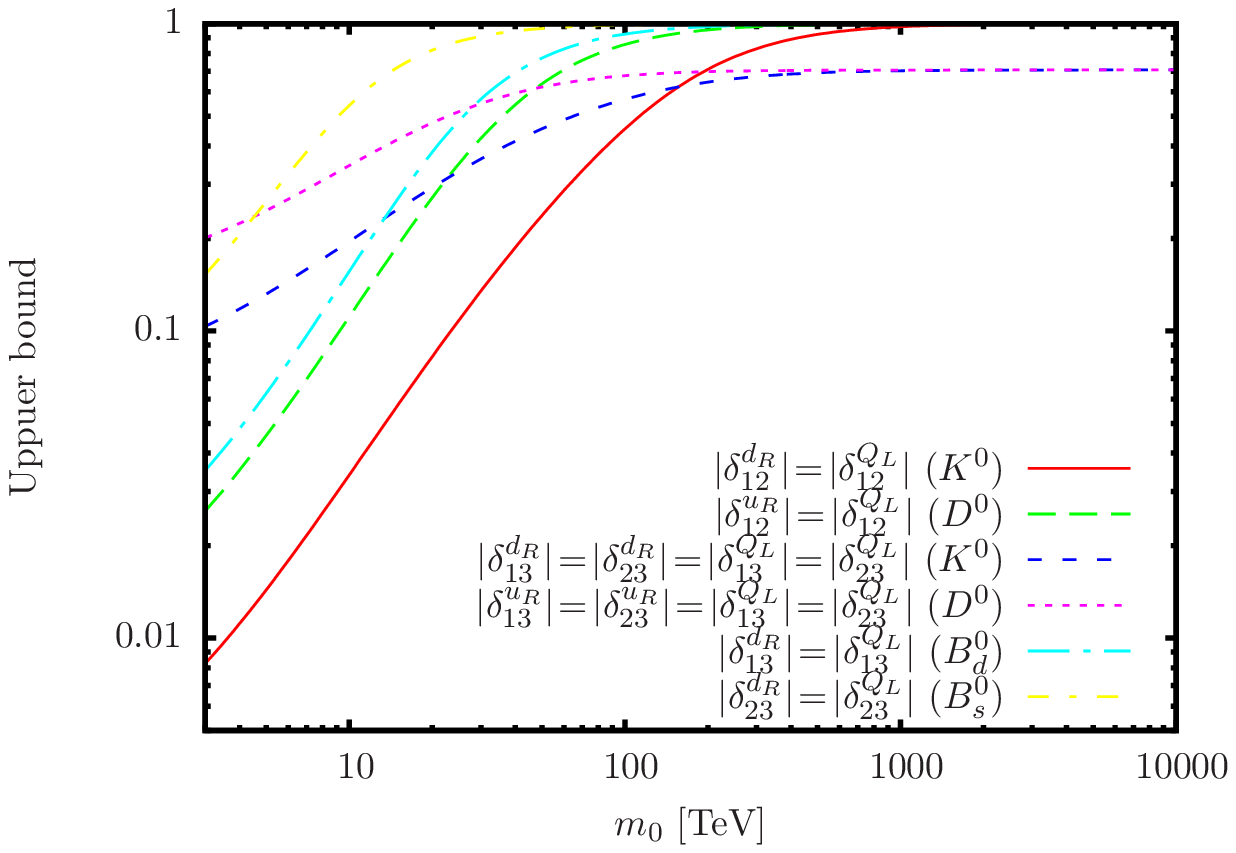}}
\subfigure[CPV + R]{\includegraphics[width=75mm]{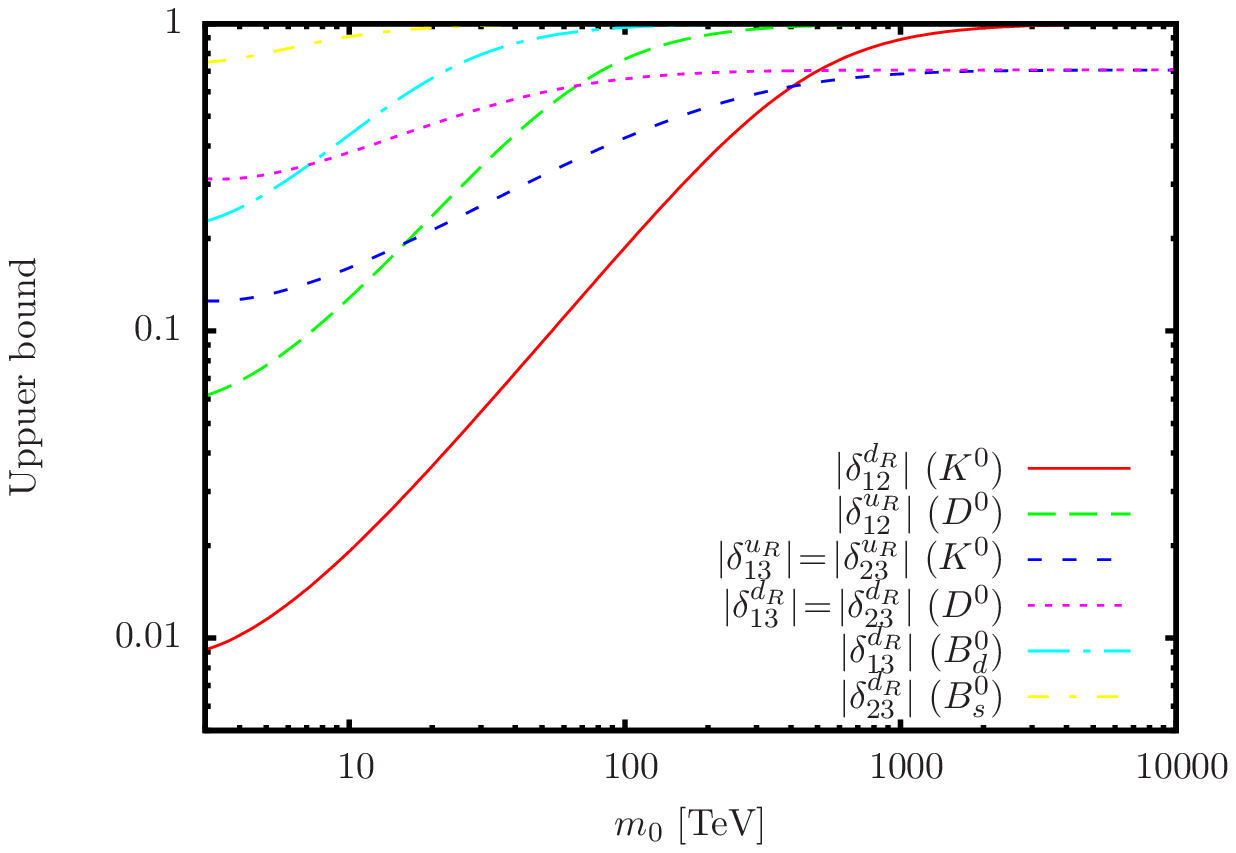}}
\subfigure[CPV + LR]{\includegraphics[width=75mm]{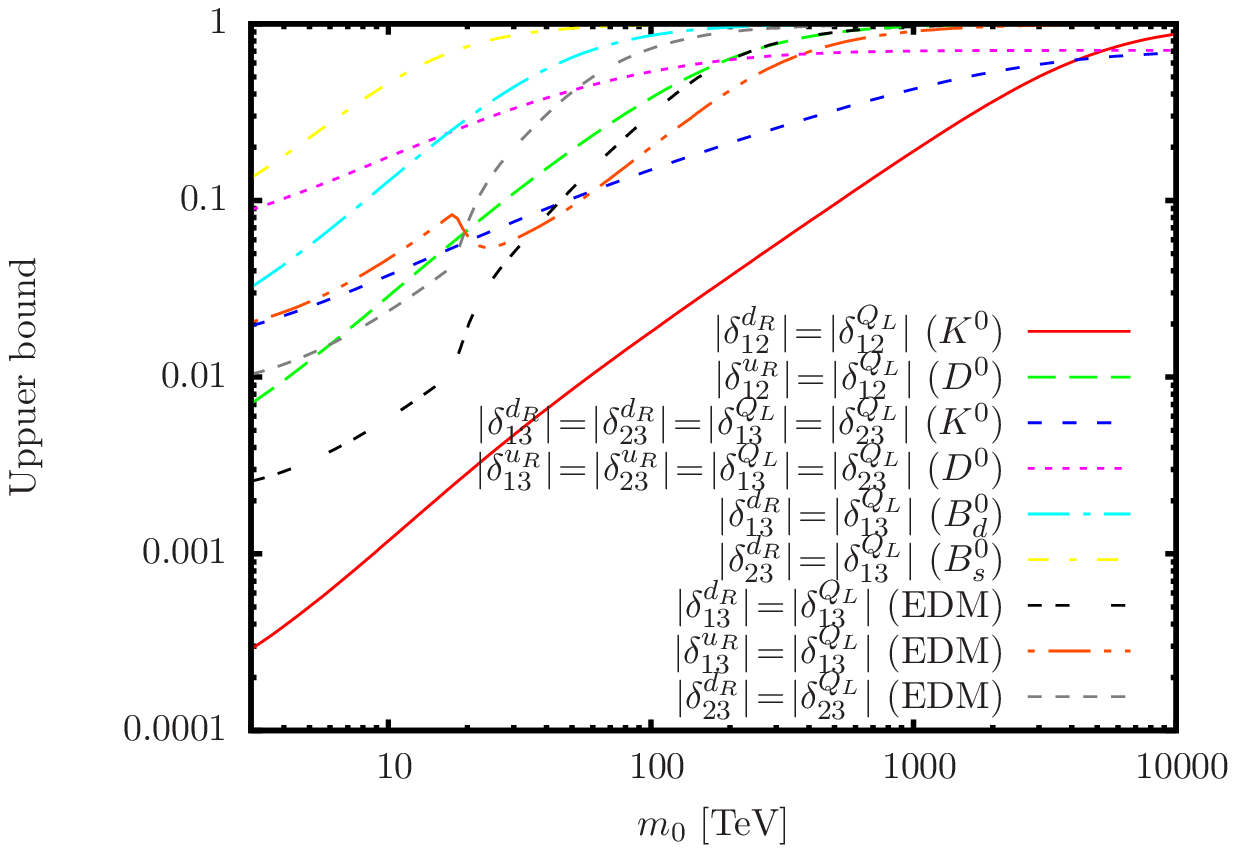}}
\caption{Upper-bound on the flavor violating mass terms $\delta$.
 See text for explanation.
}
\label{fig:constraint}
\end{center}
\end{figure}

In Fig. \ref{fig:constraint}, we show the constraints on some combinations of $\delta$'s.
$\Delta$'s and $\delta$'s which are not displayed in the figure are set to be zero.
We set the Higgsino and gauginos masses as  $\mu = m_0$, $m_{\tilde g} = -1.5$ TeV, $m_{\tilde B}=+400$ GeV $m_{\tilde W}=+ 200$ GeV.
Here we set $\tan \beta$ to realize the 125 GeV Higgs mass.
In this estimation, we neglect the flavor violation terms and assume that all the SUSY scalar particles have a common mass $m_0$ and that $m_{t} = 173.2$ GeV.
\footnote{
With large flavor violating mass terms, the mass of the lightest squark can be much less than $m_0$ and $\mu$.
In this case, large finite corrections can be expected and change the prediction of the Higgs mass.
In addition, if $|\mu| \gg m_{\rm squark}$, the electroweak breaking vacuum may become metastable or unstable.
In this paper, we neglect these effects for simplicity.
}
In this setup, for $m_0 \lsim 10 $ TeV or $m_0 \gsim 10^4$ TeV, no $\tan\beta$ for the 125 GeV Higgs mass is found 
within the range of 1 to 50. 
In such cases, we set $\tan\beta = 50$ and  $1$, respectively.
We also require the lightest squark is heavier than the gluino.

To get the constraints, we evaluate the one-loop box and dipole diagrams, 
evolve the Wilson coefficients down to relevant hadronic scales via RG equations from the QCD interaction and
estimate the hadronic matrix elements.
For the meson mixings,  we use the results of new physics fits of  Refs. \cite{Bona:2007vi,  Bevan:2012waa,UTfit} and obtain the constraints.
The constraints from $K^0-\bar{K}^0$ and $D^0-\bar{D}^0$ mixings have large uncertainties, for lack of concrete SM predictions.
In our analysis, we assume that the uncertainties of the SM predictions are same as experimental observations, e.g., $\delta(\Delta m_{K})_{\rm SM} = (\Delta m_{K})_{\rm exp}$.
Although the hadronic EDMs suffer from large uncertainties, we adopt the result of Ref. \cite{Hisano:2004tf} and
compare them with the current experimental constraints \cite{Baker:2006ts,Griffith:2009zz}.

Fig. \ref{fig:constraint}-(a) shows the case of flavor violation of right-handed squarks without CP-violation. 
Note that the constraints from $K^0-\bar{K}^0$ and $D^0-\bar{D}^0$ suffer from large ambiguities, depending on the treatment of the SM predictions.
Fig. \ref{fig:constraint}-(b) shows the case of flavor violation of both left and right-handed squarks without CP-violation.
Generally, compared with the case of (a), the constraints get severer.
Fig. \ref{fig:constraint}-(c) shows the case of flavor violation of only right-handed squarks with CP-violation.
We choose the CP-phase so that the strongest constraint can be obtained.
In this case, the constraint from  $\epsilon_K$ is strong and robust.
Fig. \ref{fig:constraint}-(d) shows the case of both left and right-handed flavor violation and  CP-violation.
In this case, we see that even PeV-scale SUSY suffers from strong constraints from the $K^0-\bar{K}^0$ mixing.
Here we show that the EDMs also give powerful constraints on the flavor violation.
However note that these constraints strongly depend on the size of $\mu$, $\tan\beta$ and hadronic uncertainties.

\subsection{Flavorful Gluino Decay}
As discussed in Section 2, the flavor violating gluino decay can be expected for the squark flavor violation.
As in Fig. \ref{fig:constraint}, if the squark mass is larger than ${\cal O}(100)$ TeV, large flavor violation is allowed from the current constraints.
Generally, there are gluino decays ($\tilde g \to q_1 q_2 \chi$) 
violating both the first and second flavor violation which comes from either minimal or non-minimal flavor violation.
However, instead,  we focus on the gluino decays accompanied with third flavor violation ($\tilde g \to q_{1,2} q_3 \chi$),
since these constraints are weak and third-family violating gluino decay may be experimentally more viable to be discovered.

\begin{figure}[htbp]
\begin{center}
\subfigure[CP + R]{\includegraphics[width=75mm]{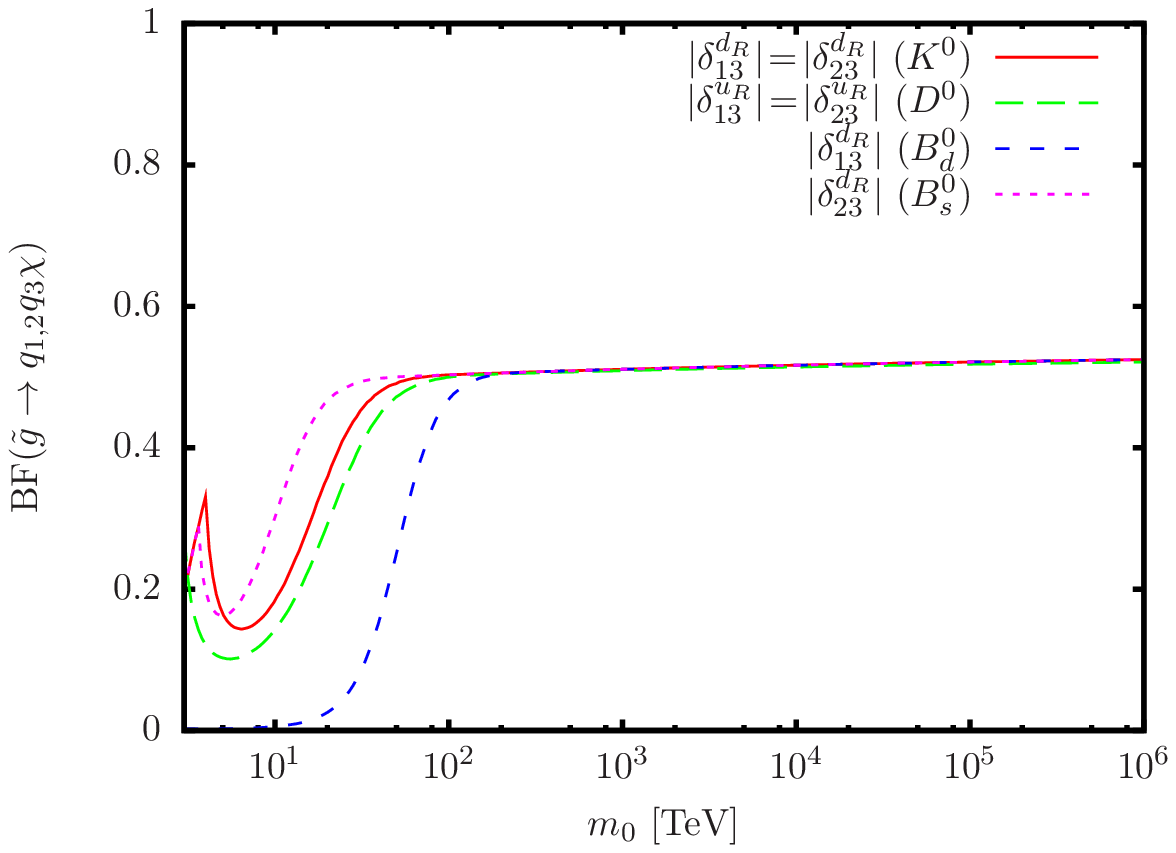}}
\subfigure[CP + LR]{\includegraphics[width=75mm]{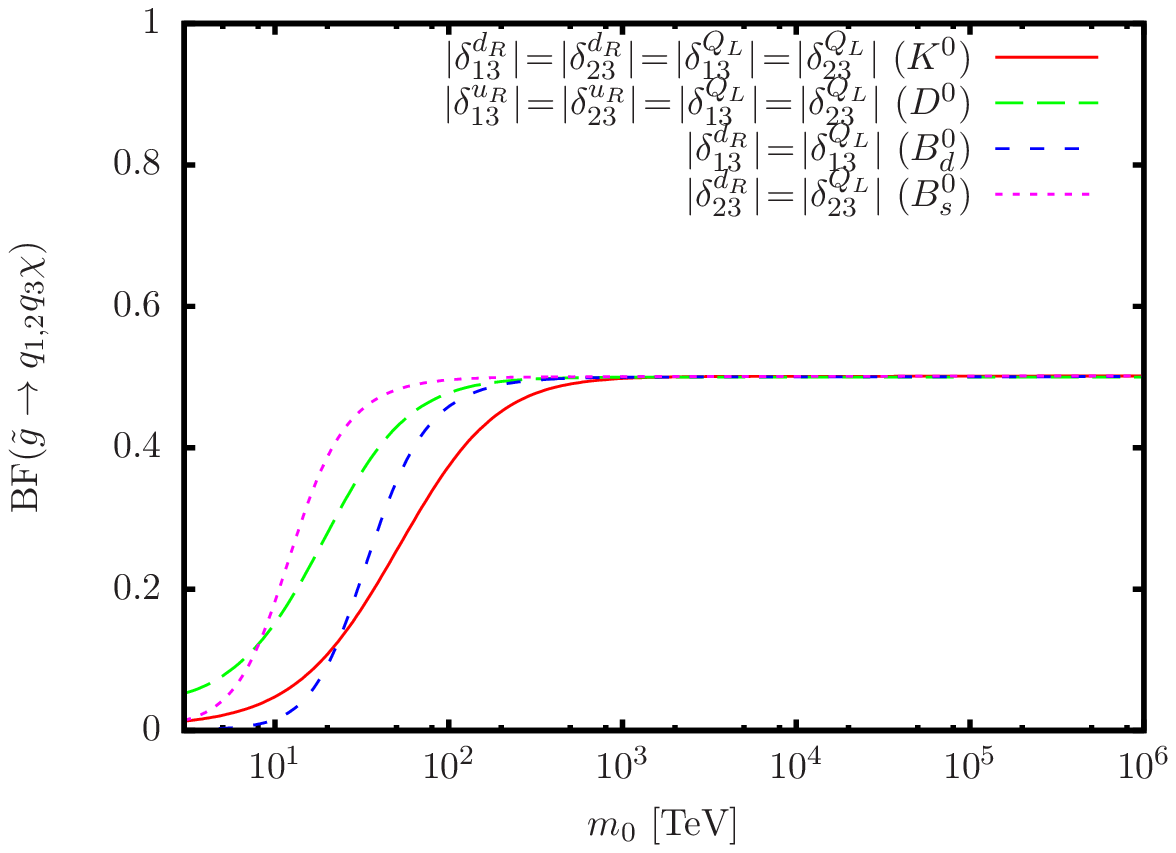}}
\subfigure[CPV + R]{\includegraphics[width=75mm]{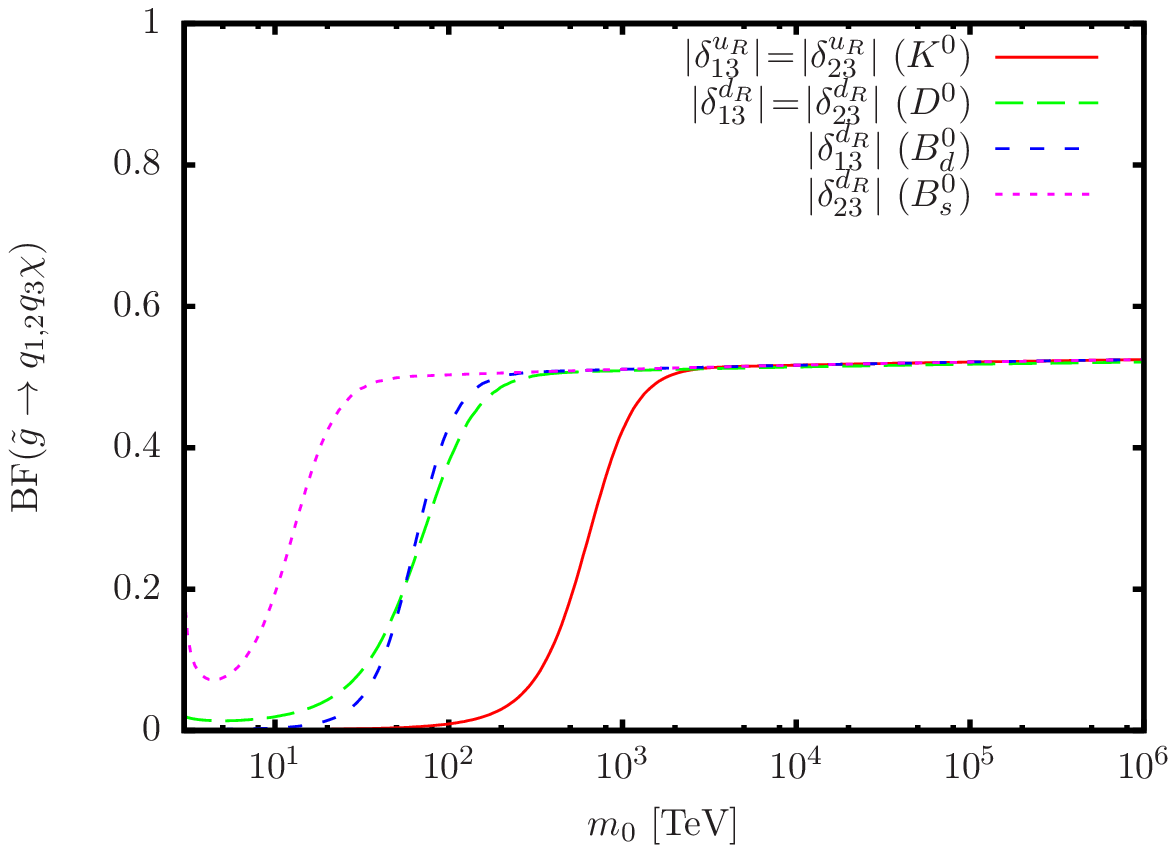}}
\subfigure[CPV + LR]{\includegraphics[width=75mm]{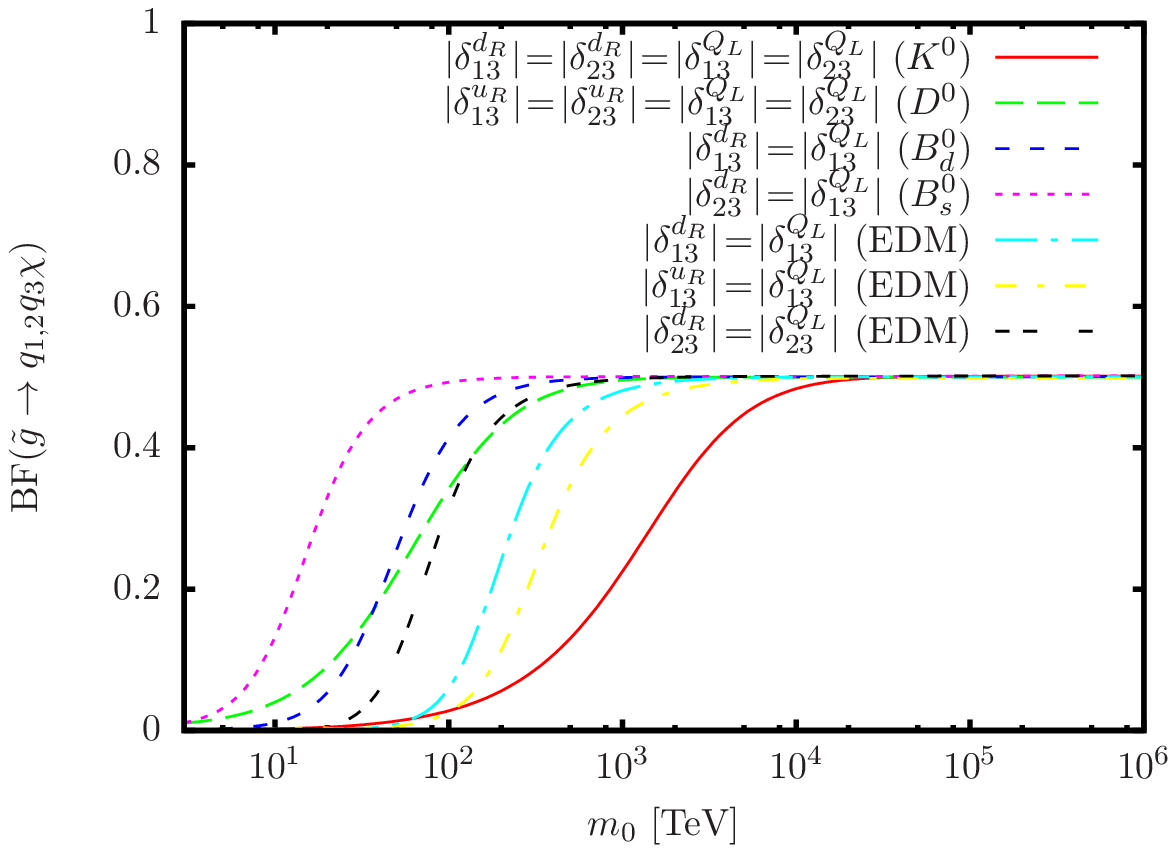}}
\caption{Upper-bound on the gluino branching fraction with third-family flavor violation.
}
\label{fig:constraint_BF}
\end{center}
\end{figure}

Using the same setup in Fig. \ref{fig:constraint}, 
we show the  upper-bound on the branching fraction of $\tilde g \to q_{1,2} q_{3} \chi$, 
keeping the current experimental constraints in Fig. \ref{fig:constraint_BF}. 
We adopt the lightest squark mass as the boundary scale $\tilde m$.
As the squark mass scale $m_0$ gets larger, the larger flavor violation $\delta$'s are allowed.
In the case of the large $\delta$'s, the lightest squarks,  which are mixed states of different generations,
dominate the gluino decay process and 
the flavor violating branching fraction approximates 50 \%.
Except for the constraints from $\epsilon_K$, the squark mass scale higher than ${\cal O}(10^{2-3})$ TeV can 
saturate the gluino flavor violating branching fraction.
As $m_0$ is larger, the RG evolution effects get more efficient and logarithmically enhance this fraction.

%%%%%%%%%%%%%%%%%%%%%%%%%%%%%%%%%%%%%%%%%%%%%%%%%%%%%%%%
% Collider
%%%%%%%%%%%%%%%%%%%%%%%%%%%%%%%%%%%%%%%%%%%%%%%%%%%%%%%%
\section{Collider Signature}
Let us briefly discuss the collider signature of the gluino decay in the high-scale SUSY models.
Distinctive features of the gluino decay in the high-scale SUSY models are the longevity of the gluino lifetime and
the flavor violating decay modes.
The signatures of the long-lived gluinos are often discussed in the high-scale SUSY scenarios and
leave unique signatures \cite{Hewett:2004nw,Cheung:2004ad}.
Instead we focus on the signals of the flavorful gluino decay.
Depending on the lifetime of the gluino, the way to identify the flavorful gluino decay differs.

\begin{description}
\item[Case I. $c\tau_{\tilde g} \ll {\cal O}(1)~{\rm mm}$  ($\tilde{m}<{\cal O}(10^3)~\TEV$)]\mbox{}\\ 
In this case, the impact parameters of the gluino decay products are less than the limitation of the LHC detector.
Therefore it is difficult to directly measure the gluino lifetime.
Instead, in this case, conventional collider techniques such as  heavy-flavor quark and lepton tagging, jet reconstructions and so on 
are available.
Depending on the gluino decay mode, distinctive signatures can be expected \cite{Bartl:2011wq}.
Using, for instance, the techniques developed  in Ref. \cite{Sato:2012xf}, 
we can estimate the gluino branching fractions.
\item[Case I\!I. ${\cal O}(1)~{\rm mm} < c\tau_{\tilde g} < {\cal O}(10)~{\rm m}$ (${\cal O}(10^3)~{\rm TeV}  <\tilde{m}<{\cal O}(10^4)~\TEV$)]\mbox{}\\ 
In this case, although the gluino can decay inside the detector, some of traditional collider techniques will fail.
For example, usage of the conventional $b$-tagging is doubtful, since it owes to impact parameters of the $b$ decays.
However, studies of collider signatures with displaced vertices are developing.
We expect the detailed analysis of track reconstructions and energy deposit in detectors can reveal the displaced gluino decays.
\item[Case I\!I\!I. $c\tau_{\tilde g} \gg {\cal O}(10) ~{\rm m}$   ($\tilde{m}\gg {\cal O}(10^4)~\TEV$)]\mbox{}\\ 
In this case, the detection of in-flight decays of the gluinos is difficult.
However, if a number of gluinos are produced, some of the gluinos can be trapped in the detector \cite{Arvanitaki:2005nq}.
Information of the energy deposit in the gluino decay may be helpful, as discussed in the context of 
measurement of the branching fraction of the trapped staus \cite{Ito:2011xs}.
For instance, $t$ quarks from the gluino decay emit leptons, which change hadronic energy deposit of the gluino decay.
Therefore a detailed study of the energy deposit of the stopped gluino decay gives us a clue for the gluino decay mode.
\end{description}
Once the gluino decays can be fully identified, ideally, ${\cal O}(10^4)$ gluino decays may result in ${\cal O}(1)$ \% accuracy of the measurement of the gluino branching fractions.
However it is unclear whether we can identify the gluino decay, eliminating the background.
The detailed analysis is beyond the scope of this paper and will be done elsewhere \cite{fv_in_gluino}.

%%%%%%%%%%%%%%%%%%%%%%%%%%%%%%%%%%%%%%%%%%%%%%%%%%%%%%%%
% Conclusion
%%%%%%%%%%%%%%%%%%%%%%%%%%%%%%%%%%%%%%%%%%%%%%%%%%%%%%%%
\section{Conclusion and Discussion}
The framework of high-scale SUSY models is well motivated from both phenomenological and theoretical viewpoints.
In particular the MSSM with the scalar scale ${\cal O}(10-10^6)$ TeV can explain the Higgs boson with a mass of about 125 GeV.
In such a framework, while the SUSY particles in the electroweak scale have a rather simple spectrum and can be accessible by experiments,
the structure of the sfermion sector has strong model dependence and cannot be directly probed by the experiments.
Examining the sfermion sector is essential to discriminate the underlying model.
In this paper, we have discussed the gluino decay as a probe of the sfermion sector in a high-scale SUSY model.
We have studied the relation between the squark sector and the gluino decay in detail.

If the gaugino mass scale is within the reach of a collider,
we can observe the gluino decay which is strongly affected by the squark sector.
The gluino lifetime reflects the squark scale and  the decay products are sensitive to the squark flavor structure.
Therefore the gluino decay will provide us clues to investigate the heavy sfermion sector.
Especially, in our study, we focused on the flavor structure of the gluino decay and 
investigated the correlation with low-energy observations.

The large mass hierarchy between the sfermion and gaugino scale can result in large quantum corrections for the gluino decay.
We have used RG methods to tame the quantum corrections.
We have found that the squark mass scale is higher than ${\cal O}(100)$ TeV, the quantum corrections can affect the branching ratio by ${\cal O}(10)$ \%, compared to 
the tree-level results.
Ideally, by observing ${\cal O}(10^4)$ gluino decays at a collider, we can estimate the branching fraction with ${\cal O}(1)$ \% accuracy.
Compared to this accuracy, the effects of the quantum corrections are more significant, 
when we try to reconstruct the squark sector via measurement of the gluino decays.

We also discussed the correlation between the gluino decay and low-energy flavor/CP observations.
From the current constraints from the flavor/CP, we have found that  the scalar scale ${\cal O}(10-10^6)$ TeV allows large flavor violating gluino decay.
Unlike the flavor structure of the gluino decay products, measurement of CP violation of the gluino decay seems hard.
On the other hand, (future) low-energy flavor/CP experiments are sensitive to such CP violation.
The combination of the gluino decay, the lifetime and decay modes, and  low-energy flavor/CP observations is crucially important for
the study of the underlying flavor/CP symmetry of SSMs.

In this study, we focus on the flavor structure of the quark sector.
Recently, flavor violation of a leptonic sector in a high-scale SUSY model is also studied \cite{Moroi:2013sfa}.
Combining these sfermion information will provide us a hint for an underlying flavor and GUT model.

\section*{Acknowledgements}
SS thanks L. Hall and Y. Nomura for fruitful discussion.
The work of RS and KT is supported in part by JSPS Research Fellowships for Young Scientists.
This work is also supported by the World Premier International Research Center Initiative (WPI Initiative), MEXT, Japan.

%%%%%%%%%%%%%%%%%%%%%%%%%%%%%%%%%%%%%%%%%%%%%%%%%%%%%%%%
% appendix
%%%%%%%%%%%%%%%%%%%%%%%%%%%%%%%%%%%%%%%%%%%%%%%%%%%%%%%%
\appendix

\section{Note on Flavor Basis}\label{sec:notation}
In this Appendix, we summarize notations of the flavor structure which is used in this paper.
We neglect $A$-terms of squarks, then, the flavor dependent terms in the Lagrangian is written as,
\begin{align}
{\cal L}_{\rm Yukawa} + {\cal L}_{\rm soft} &= (H Q_L Y_u u_R^c + H^* Q_L Y_d d_R^c + h.c.) 
 - {\tilde Q_L}^* m^2_{\tilde Q_L}{\tilde Q_L} - {\tilde u_R}^{c} m^2_{\tilde u_R}{\tilde u_R^{c*}} - {\tilde d_R}^{c} m^2_{\tilde d_R}{\tilde d_R^{c*}},
\end{align}

\subsection{Weak Basis}
We can redefine quark and squark fields by using $SU(2)_L \times U(1)_Y$ symmetric redefinition.
\begin{align}
Q_{L,i} \to Q'_{L,i} = U_{ij}^Q Q_{L,j},\qquad &\tilde Q_{L,i} \to \tilde Q'_{L,i} = U_{ij}^Q \tilde Q_{L,j},\\
u_{R,i}^c \to {u_{R,i}^c}' = U_{ij}^u u_{L,j}^c,\qquad &\tilde u_{R,i}^c \to \tilde {u_{R,i}^c}' = U_{ij}^u \tilde u_{R,j}^c,\\
d_{R,i}^c \to {d_{R,i}^c}' = U_{ij}^d d_{L,j}^c,\qquad &\tilde d_{R,i}^c \to \tilde {d_{R,i}^c}' = U_{ij}^d \tilde d_{R,j}^c.
\end{align}
By using the above redefinition, we take a basis in which Yukawa matrices at electroweak scale is given as, \begin{eqnarray}
Y_u = V_{\rm CKM}^T {\rm diag}(y_u, y_c, y_t), \qquad
Y_d = {\rm diag}(y_d, y_s, y_b).
\end{eqnarray}
Note on that RG evolution of Yukawa matrices destroys the above form except for electroweak scale because we take scale independent $U$'s.
We denote this basis as ``weak basis'' in this paper.
In this basis, we can write interaction terms as $SU(2)_L$ symmetric forms,
then, this basis is convenient to calculate the RG equations.

\subsection{Super-CKM basis}
In this basis, Yukawa interaction of quarks are diagonalized at electroweak scale.
Quarks and squarks in this basis are related to ones in weak-basis as,
\begin{eqnarray}
\hat{u}_L = V_{\rm CKM} u_L,\qquad
\hat{u}_R^c = u_R^c,\qquad
\hat{d}_L = d_L,\qquad
\hat{d}_R^c = d_R^c,\\
\hat{\tilde u}_L = V_{\rm CKM} \tilde u_L,\qquad
\hat{\tilde u}_R^c = \tilde u_R^c,\qquad
\hat{\tilde d}_L = \tilde d_L,\qquad
\hat{\tilde d}_R^c = \tilde d_R^c.
\end{eqnarray}
Each quark is mass eigenstate, and each squark is defined as a super-partner of each quark.
In this basis, mass-squared matrices of squarks are given by,
\begin{align}
{\cal L}_{\rm soft} &= 
- \hat{\tilde{u}}_L^* \hat{m}^2_{\tilde{u}_L} \hat{\tilde{u}}_L
- \hat{\tilde{d}}_L^* \hat{m}^2_{\tilde{d}_L} \hat{\tilde{d}}_L
- \hat{\tilde{u}}_R^{c} \hat{m}^2_{\tilde{u}_R} \hat{\tilde{u}}_R^{c*}
- \hat{\tilde{d}}_R^{c} \hat{m}^2_{\tilde{d}_R} \hat{\tilde{d}}_R^{c*},
\end{align}
where, 
\begin{align}
\hat{m}_{\tilde{u}_L}^2 = V_{\rm CKM} m_{\tilde{Q}_L}^2 V_{\rm CKM}^\dagger,\qquad
\hat{m}_{\tilde{d}_L}^2 = m_{\tilde{Q}_L}^2,\qquad
\hat{m}_{\tilde{u}_R}^2 = m_{\tilde{u}_R}^2,\qquad
\hat{m}_{\tilde{d}_R}^2 = m_{\tilde{d}_R}^2.
\end{align}

\section{RG Equations for Wilson Coefficients of Higgsinos}\label{sec:higgsinoRG}
In some high-scale SUSY models, the Higgsino mass is also much smaller than the sfermion and heavy Higgs boson masses.
If the gluino is heavier than the Higgsinos, it can decay into the Higgsino.
Such decay processes are induced by the following dimension six effective interactions:
\begin{eqnarray}
Q^{\tilde H}_{1u,ij} &=& (\tilde H_u \tilde g)(Q_{L,i} u_{R,j}^c), \\
Q^{\tilde H}_{2u,ij} &=& (\tilde H_u \s^{\m\n} \tilde g)(Q_{L,i} \s_{\m\n} u_{R,j}^c), \\
Q^{\tilde H}_{5u} &=& (\tilde H_u \s^{\m\n} \tilde g) H G_{\m\n}, \\
Q^{\tilde H}_{1d,ij} &=& (\tilde H_d \tilde g)(Q_{L,i} d_{R,j}^c), \\
Q^{\tilde H}_{2d,ij} &=& (\tilde H_d \s^{\m\n} \tilde g)(Q_{L,i} \s_{\m\n} d_{R,j}^c), \\
Q^{\tilde H}_{5d} &=& (\tilde H_d \s^{\m\n} \tilde g) H^* G_{\m\n}.
\end{eqnarray}
Dimensionless couplings in MSSM respect PQ-symmetry, 
then, at the leading order, we do not have to consider PQ-breaking operators, e.g., $(\tilde H_d \tilde g)(q_{L,i} u_{R,j}^c)^\dagger$.
The effective Lagrangian contains the following interaction terms:
{\small
\begin{eqnarray}
{\cal L}_{\rm eff} &=&
\frac{1}{\tilde m^2} \Biggl[
\sum_{i,j}
\left(
C^{\tilde H}_{1u,ij} Q^{\tilde H}_{1u,ij}
+ C^{\tilde H}_{2u,ij} Q^{\tilde H}_{2u,ij}
+ C^{\tilde H}_{1d,ij} Q^{\tilde H}_{1d,ij}
+ C^{\tilde H}_{2d,ij} Q^{\tilde H}_{2d,ij}
\right) \nonumber\\
&& \quad\quad\quad
+ C^{\tilde H}_{5u} Q^{\tilde H}_{5u}
+ C^{\tilde H}_{5d} Q^{\tilde H}_{5d} \Biggr] + h.c..
\end{eqnarray}
}
The boundary conditions at the squark mass scale $\tilde m$ are given by,
\begin{eqnarray}
C^{\tilde H}_{1u}(\tilde m) &=& \frac{g_s}{\sqrt{2}\sin\b} \tilde m^2 \left[ (m_{\tilde Q_L}^2)^{-1} Y_u - Y_u ((m_{\tilde u_R}^2)^{-1})^T \right],\\
C^{\tilde H}_{2u}(\tilde m) &=& \frac{g_s}{4\sqrt{2}\sin\b} \tilde m^2 \left[ (m_{\tilde Q_L}^2)^{-1} Y_u + Y_u ((m_{\tilde u_R}^2)^{-1})^T \right],\\
C^{\tilde H}_{5u}(\tilde m) &=& \frac{g_s^2}{32\sqrt{2} \pi^2 \sin\b} \tilde m^2 \left( {\rm tr}[ Y_u Y_u^\dagger (m_{\tilde Q_L}^2)^{-1} ] +  {\rm tr}[ Y_u^* Y_u^T (m_{\tilde u_R}^2)^{-1} ] \right),\\
C^{\tilde H}_{1d}(\tilde m) &=& \frac{g_s}{\sqrt{2}\cos\b} \tilde m^2 \left[ (m_{\tilde Q_L}^2)^{-1} Y_d - Y_d ((m_{\tilde d_R}^2)^{-1})^T \right],\\
C^{\tilde H}_{2d}(\tilde m) &=& -\frac{g_s}{4\sqrt{2}\cos\b} \tilde m^2 \left[ (m_{\tilde Q_L}^2)^{-1} Y_d + Y_d ((m_{\tilde d_R}^2)^{-1})^T \right],\\
C^{\tilde H}_{5d}(\tilde m) &=& \frac{g_s^2}{32\sqrt{2} \pi^2 \cos\b} \tilde m^2 \left( {\rm tr}[ Y_d Y_d^\dagger (m_{\tilde Q_L}^2)^{-1} ] +  {\rm tr}[ Y_d^* Y_d^T (m_{\tilde d_R}^2)^{-1} ] \right).
\end{eqnarray}
The RG equations for the Wilson coefficients $C^{\tilde H}$'s are given by,
{\small
\begin{eqnarray}
16\pi^2 \frac{d C^{\tilde H}_{1u}}{d\log\m} &=&
	\frac{3g_s^2}{N_C} C^{\tilde H}_{1u}
	+ \frac{1}{2} (Y_u Y_u^\dagger + Y_d Y_d^\dagger) C^{\tilde H}_{1u}
	+ C^{\tilde H}_{1u} Y_u^\dagger Y_u \label{eq:RGE_higgsino_1u}
, \\
16\pi^2 \frac{d C^{\tilde H}_{2u}}{d\log\m} &=&
	-\left( 4N_C + \frac{1}{N_C} \right) g_s^2 C^{\tilde H}_{2u}
	+ \frac{1}{2} (Y_u Y_u^\dagger + Y_d Y_d^\dagger) C^{\tilde H}_{2u}
	+ C^{\tilde H}_{2u} Y_u^\dagger Y_u \label{eq:RGE_higgsino_2u}
	+ 2 g_s Y_u C^{\tilde H}_{5u}
, \\
16\pi^2 \frac{d C^{\tilde H}_{5u}}{d\log\m}
	&=& \left( \frac{2}{3}N_F - 6 N_C \right) g_s^2 C^{\tilde H}_{5u}
	+ N_C {\rm tr} \left[ Y_u Y_u^\dagger + Y_d Y_d^\dagger \right] C^{\tilde H}_{5u}
	+ 4 g_s {\rm tr}\left[ C_{2u} Y_u^\dagger \right]\label{eq:RGE_higgsino_5u}
, \\
16\pi^2 \frac{d C^{\tilde H}_{1d}}{d\log\m} &=&
	\frac{3g_s^2}{N_C} C^{\tilde H}_{1d}
	+ \frac{1}{2} (Y_u Y_u^\dagger + Y_d Y_d^\dagger) C^{\tilde H}_{1d}
	+ C^{\tilde H}_{1d} Y_d^\dagger Y_d\label{eq:RGE_higgsino_1d}
, \\
16\pi^2 \frac{d C^{\tilde H}_{2d}}{d\log\m} &=&
	-\left( 4N_C + \frac{1}{N_C} \right) g_s^2 C^{\tilde H}_{2d}
	+ \frac{1}{2} (Y_u Y_u^\dagger + Y_d Y_d^\dagger) C^{\tilde H}_{2d}
	+ C^{\tilde H}_{2d} Y_d^\dagger Y_d\label{eq:RGE_higgsino_2d}
	+ 2 g_s Y_d C^{\tilde H}_{5d}
, \\
16\pi^2 \frac{d C^{\tilde H}_{5d}}{d\log\m}
	&=& \left( \frac{2}{3}N_F - 6 N_C \right) g_s^2 C^{\tilde H}_{5d}
	+ N_C {\rm tr} \left[ Y_u Y_u^\dagger + Y_d Y_d^\dagger \right] C^{\tilde H}_{5d}
	+ 4 g_s {\rm tr}\left[ C_{2d} Y_d^\dagger \right]\label{eq:RGE_higgsino_5d}
.
\end{eqnarray}
}
Here, we take into account only quark Yukawa and QCD interactions.
We have checked Eqs. (\ref{eq:RGE_higgsino_1u}--\ref{eq:RGE_higgsino_5d}) are consistent with Ref. \cite{Gambino:2005eh} in the flavor-symmetric case.
For the RG equations of dimensionless coupling constants, see the Appendix of Ref. \cite{Giudice:2011cg}.

\section{RG Equations of Soft Mass with Flavor Violation}
In this appendix, we show the RG equations for sfermion soft masses.
Here, we neglect trilinear-coupling, gaugino masses.
We assume relatively small $\tan\b$, then neglect down-type quark and lepton Yukawa couplings.
{\small
\begin{eqnarray}
16\pi^2 \frac{d m_{\tilde Q_L}^2}{d\log\m} &=& 2 \left\{ Y_u^* Y_u^T , m_{\tilde Q_L}^2 \right\} + 2 Y_u^* (m_{\tilde u_R}^{2T} + m_{H_u}^2 {\bf 1})Y_u^T + \frac{g_1^2 S}{5}{\bf 1},\\
16\pi^2 \frac{d m_{\tilde u_R}^2}{d\log\m} &=& 4 \left\{ Y_u^\dagger Y_u, m_{\tilde u_R}^2 \right\} + 4 Y_u^\dagger (m_{\tilde Q_L}^{2T} + m_{H_u}^2 {\bf 1})Y_u - \frac{4 g_1^2 S}{5}{\bf 1},\\
16\pi^2 \frac{d m_{\tilde d_R}^2}{d\log\m} &=& \frac{2 g_1^2 S}{5}{\bf 1},\\
16\pi^2 \frac{d m_{\tilde L_L}^2}{d\log\m} &=& - \frac{3 g_1^2 S}{5}{\bf 1},\\
16\pi^2 \frac{d m_{\tilde e_R}^2}{d\log\m} &=& \frac{6 g_1^2 S}{5}{\bf 1},\\
16\pi^2 \frac{d m_{H_u}^2}{d\log\m} &=& 6 {\rm tr}[ Y_u^* Y_u^T m_{\tilde Q_L}^2 ] + 6 {\rm tr} [ Y_u^\dagger Y_u m_{\tilde u_R}^2 ] + 6 {\rm tr}[ Y_u Y_u^\dagger ] m_{H_u}^2 + \frac{3 g_1^2 S}{5},\\
16\pi^2 \frac{d m_{H_d}^2}{d\log\m} &=& - \frac{3 g_1^2 S}{5}.
 \end{eqnarray}
}
Here, sfermion mass-squared matrices $m_{\tilde Q_L}^2$, $m_{\tilde u_R}^2$, $m_{\tilde d_R}^2$, $m_{\tilde L_L}^2$ and $m_{\tilde e_R}^2$ and Yukawa matrix $Y_u$ are $3\times 3$ matrices in flavor space.
$S  = m_{H_u}^2 - m_{H_d}^2  + {\rm tr}\left[ m_Q^2 - 2 m_u^2 + m_d^2 - m_L^2 + m_e^2 \right]$.
$\{ A,B \} = AB + BA$, ${\bf 1} = {\rm diag}(1,1,1)$.

%%%%%%%%%%%%%%%%%%%%%%%%%%%%%%%%%%%%%%%%%%%%%%%%%%%%%%%%
% references
%%%%%%%%%%%%%%%%%%%%%%%%%%%%%%%%%%%%%%%%%%%%%%%%%%%%%%%%


\begin{thebibliography}{99}
%\cite{Aad:2012tfa}
\bibitem{Aad:2012tfa} 
  G.~Aad {\it et al.}  [ATLAS Collaboration],
  %``Observation of a new particle in the search for the Standard Model Higgs boson with the ATLAS detector at the LHC,''
  Phys.\ Lett.\ B {\bf 716}, 1 (2012)
  [arXiv:1207.7214 [hep-ex]].
  %%CITATION = ARXIV:1207.7214;%%
  %1309 citations counted in INSPIRE as of 05 Jul 2013
%\cite{Chatrchyan:2012ufa}
\bibitem{Chatrchyan:2012ufa} 
  S.~Chatrchyan {\it et al.}  [CMS Collaboration],
  %``Observation of a new boson at a mass of 125 GeV with the CMS experiment at the LHC,''
  Phys.\ Lett.\ B {\bf 716}, 30 (2012)
  [arXiv:1207.7235 [hep-ex]].
  %%CITATION = ARXIV:1207.7235;%%
  %1295 citations counted in INSPIRE as of 05 Jul 2013


%\cite{Wells:2003tf}
\bibitem{Wells:2003tf} 
  J.~D.~Wells,
  %``Implications of supersymmetry breaking with a little hierarchy between gauginos and scalars,''
  hep-ph/0306127.
  %%CITATION = HEP-PH/0306127;%%
  %49 citations counted in INSPIRE as of 30 May 2013
  
  

  

  %\cite{ArkaniHamed:2004fb}
\bibitem{ArkaniHamed:2004fb}
  N.~Arkani-Hamed and S.~Dimopoulos,
  %``Supersymmetric unification without low energy supersymmetry and signatures for fine-tuning at the LHC,''
  JHEP {\bf 0506}, 073 (2005)
  [hep-th/0405159].
  %%CITATION = HEP-TH/0405159;%%
%\cite{Giudice:2004tc}
\bibitem{Giudice:2004tc}
  G.~F.~Giudice and A.~Romanino,
  %``Split supersymmetry,''
  Nucl.\ Phys.\ B {\bf 699}, 65 (2004)
  [Erratum-ibid.\ B {\bf 706}, 65 (2005)]
  [hep-ph/0406088].
  %%CITATION = HEP-PH/0406088;%%
  %\cite{ArkaniHamed:2004yi}
\bibitem{ArkaniHamed:2004yi}
  N.~Arkani-Hamed, S.~Dimopoulos, G.~F.~Giudice and A.~Romanino,
  %``Aspects of split supersymmetry,''
  Nucl.\ Phys.\ B {\bf 709}, 3 (2005)
  [hep-ph/0409232].
  %%CITATION = HEP-PH/0409232;%%

  %\cite{Wells:2004di}
\bibitem{Wells:2004di} 
  J.~D.~Wells,
  %``PeV-scale supersymmetry,''
  Phys.\ Rev.\ D {\bf 71}, 015013 (2005)
  [hep-ph/0411041].
  %%CITATION = HEP-PH/0411041;%%
  %112 citations counted in INSPIRE as of 11 Jun 2013
  
  


\bibitem{OYY} 
%\cite{Okada:1990vk}
%\bibitem{Okada:1990vk} 
  Y.~Okada, M.~Yamaguchi and T.~Yanagida,
  %``Upper bound of the lightest Higgs boson mass in the minimal supersymmetric standard model,''
  Prog.\ Theor.\ Phys.\  {\bf 85}, 1 (1991);
  %%CITATION = PTPKA,85,1;%%
  %1026 citations counted in INSPIRE as of 12 Apr 2013
%\cite{Okada:1990gg}
%\bibitem{Okada:1990gg} 
  Y.~Okada, M.~Yamaguchi and T.~Yanagida,
  %``Renormalization group analysis on the Higgs mass in the softly broken supersymmetric standard model,''
  Phys.\ Lett.\ B {\bf 262}, 54 (1991);
  %%CITATION = PHLTA,B262,54;%%
  %471 citations counted in INSPIRE as of 12 Apr 2013
%\cite{Ellis:1990nz}
%\bibitem{Ellis:1990nz} 
  J.~R.~Ellis, G.~Ridolfi and F.~Zwirner,
  %``Radiative corrections to the masses of supersymmetric Higgs bosons,''
  Phys.\ Lett.\ B {\bf 257}, 83 (1991);
  %%CITATION = PHLTA,B257,83;%%
  %1133 citations counted in INSPIRE as of 12 Apr 2013
%\cite{Haber:1990aw}
%\bibitem{Haber:1990aw} 
  H.~E.~Haber and R.~Hempfling,
  %``Can the mass of the lightest Higgs boson of the minimal supersymmetric model be larger than m(Z)?,''
  Phys.\ Rev.\ Lett.\  {\bf 66}, 1815 (1991);
  %%CITATION = PRLTA,66,1815;%%
%\cite{Haber:1990aw}
%\bibitem{Haber:1990aw} 
  H.~E.~Haber and R.~Hempfling,
  %``Can the mass of the lightest Higgs boson of the minimal supersymmetric model be larger than m(Z)?,''
  Phys.\ Rev.\ Lett.\  {\bf 66}, 1815 (1991);
  %%CITATION = PRLTA,66,1815;%%
  %1115 citations counted in INSPIRE as of 12 Apr 2013
%\cite{Ellis:1991zd}
%\bibitem{Ellis:1991zd} 
  J.~R.~Ellis, G.~Ridolfi and F.~Zwirner,
  %``On radiative corrections to supersymmetric Higgs boson masses and their implications for LEP searches,''
  Phys.\ Lett.\ B {\bf 262}, 477 (1991).
  %%CITATION = PHLTA,B262,477;%%
  %633 citations counted in INSPIRE as of 12 Apr 2013
  
  
  
%\cite{Giudice:2011cg}
\bibitem{Giudice:2011cg} 
  G.~F.~Giudice and A.~Strumia,
  %``Probing High-Scale and Split Supersymmetry with Higgs Mass Measurements,''
  Nucl.\ Phys.\ B {\bf 858}, 63 (2012)
  [arXiv:1108.6077 [hep-ph]].
  %%CITATION = ARXIV:1108.6077;%%
  %77 citations counted in INSPIRE as of 11 Jun 2013
  
  
  
 \bibitem{AMSB}
%\cite{Giudice:1998xp}
%\bibitem{Giudice:1998xp} 
  G.~F.~Giudice, M.~A.~Luty, H.~Murayama and R.~Rattazzi,
  %``Gaugino mass without singlets,''
  JHEP {\bf 9812}, 027 (1998)
  [hep-ph/9810442];
  %%CITATION = HEP-PH/9810442;%%
  %898 citations counted in INSPIRE as of 07 Apr 2013
%\cite{Randall:1998uk}
%\bibitem{Randall:1998uk} 
  L.~Randall and R.~Sundrum,
  %``Out of this world supersymmetry breaking,''
  Nucl.\ Phys.\ B {\bf 557}, 79 (1999)
  [hep-th/9810155];
  %%CITATION = HEP-TH/9810155;%%
  %1161 citations counted in INSPIRE as of 07 Apr 2013
%\cite{Dine:1992yw}
%\bibitem{Dine:1992yw} 
  M.~Dine and D.~MacIntire,
  %``Supersymmetry, naturalness, and dynamical supersymmetry breaking,''
  Phys.\ Rev.\ D {\bf 46}, 2594 (1992)
  [hep-ph/9205227];
  %%CITATION = HEP-PH/9205227;%%
  %57 citations counted in INSPIRE as of 07 Apr 2013
%\cite{Bagger:1999rd}
%\bibitem{Bagger:1999rd} 
  J.~A.~Bagger, T.~Moroi and E.~Poppitz,
  %``Anomaly mediation in supergravity theories,''
  JHEP {\bf 0004}, 009 (2000)
  [hep-th/9911029];
  %%CITATION = HEP-TH/9911029;%%
  %212 citations counted in INSPIRE as of 07 Apr 2013
%\cite{Binetruy:2000md}
%\bibitem{Binetruy:2000md} 
  P.~Binetruy, M.~K.~Gaillard and B.~D.~Nelson,
  %``One loop soft supersymmetry breaking terms in superstring effective theories,''
  Nucl.\ Phys.\ B {\bf 604}, 32 (2001)
  [hep-ph/0011081].
  %%CITATION = HEP-PH/0011081;%%
  %86 citations counted in INSPIRE as of 07 Apr 2013
  
  
  \bibitem{spread}
%\cite{Hall:2011jd}
%\bibitem{Hall:2011jd} 
  L.~J.~Hall and Y.~Nomura,
  %``Spread Supersymmetry,''
  JHEP {\bf 1201}, 082 (2012)
  [arXiv:1111.4519 [hep-ph]];
  %%CITATION = ARXIV:1111.4519;%%
  %26 citations counted in INSPIRE as of 12 Apr 2013
%\cite{Hall:2012zp}
%\bibitem{Hall:2012zp}
  L.~J.~Hall, Y.~Nomura and S.~Shirai,
  %``Spread Supersymmetry with Wino LSP: Gluino and Dark Matter Signals,''
  JHEP {\bf 1301}, 036 (2013)
  [arXiv:1210.2395 [hep-ph]].
  %%CITATION = ARXIV:1210.2395;%%
  %19 citations counted in INSPIRE as of 12 Apr 2013

  
  
\bibitem{PGMs}
%\bibitem{Ibe:2011aa} 
  M.~Ibe and T.~T.~Yanagida,
  %``The Lightest Higgs Boson Mass in Pure Gravity Mediation Model,''
  Phys.\ Lett.\ B {\bf 709}, 374 (2012)
  [arXiv:1112.2462 [hep-ph]];
  %%CITATION = ARXIV:1112.2462;%%
  %43 citations counted in INSPIRE as of 20 Mar 2013
%\cite{Ibe:2012hu}
%\bibitem{Ibe:2012hu} 
  M.~Ibe, S.~Matsumoto and T.~T.~Yanagida,
  %``Pure Gravity Mediation with m_{3/2} = 10-100TeV,''
  Phys.\ Rev.\ D {\bf 85}, 095011 (2012)
  [arXiv:1202.2253 [hep-ph]].
  %%CITATION = ARXIV:1202.2253;%%
  %32 citations counted in INSPIRE as of 20 Mar 2013
  
   
   
   
   %\cite{Arvanitaki:2012ps}
\bibitem{Arvanitaki:2012ps} 
  A.~Arvanitaki, N.~Craig, S.~Dimopoulos and G.~Villadoro,
  %``Mini-Split,''
  JHEP {\bf 1302}, 126 (2013)
  [arXiv:1210.0555 [hep-ph]].
  %%CITATION = ARXIV:1210.0555;%%
  %24 citations counted in INSPIRE as of 30 May 2013


%\cite{ArkaniHamed:2012gw}
\bibitem{ArkaniHamed:2012gw} 
  N.~Arkani-Hamed, A.~Gupta, D.~E.~Kaplan, N.~Weiner and T.~Zorawski,
  %``Simply Unnatural Supersymmetry,''
  arXiv:1212.6971 [hep-ph].
  %%CITATION = ARXIV:1212.6971;%%
  %19 citations counted in INSPIRE as of 30 May 2013




\bibitem{Ibe:2006de}
M.~Ibe, T.~Moroi and T.~T.~Yanagida,
%``Possible Signals of Wino LSP at the Large Hadron Collider,''
Phys.\ Lett.\ B {\bf 644}, 355 (2007)
[hep-ph/0610277];
%%CITATION = HEP-PH/0610277;%%
%\bibitem{Buckley:2009kv}
M.~R.~Buckley, L.~Randall and B.~Shuve,
%``LHC Searches for Non-Chiral Weakly Charged Multiplets,''
JHEP {\bf 05}, 097 (2011)
[arXiv:0909.4549 [hep-ph]].
%%CITATION = ARXIV:0909.4549;%%

\bibitem{Asai:2007sw}
S.~Asai, T.~Moroi, K.~Nishihara and T.~T.~Yanagida,
%``Testing the Anomaly Mediation at the LHC,''
Phys.\ Lett.\ B {\bf 653}, 81 (2007)
[arXiv:0705.3086 [hep-ph]];
%%CITATION = ARXIV:0705.3086;%%
%\bibitem{Asai:2008sk}
S.~Asai, T.~Moroi and T.~T.~Yanagida,
%``Test of Anomaly Mediation at the LHC,''
Phys.\ Lett.\ B {\bf 664}, 185 (2008)
[arXiv:0802.3725 [hep-ph]];
%%CITATION = ARXIV:0802.3725;%%
%\bibitem{Asai:2008im}
S.~Asai, Y.~Azuma, O.~Jinnouchi, T.~Moroi, S.~Shirai and T.~T.~Yanagida,
%``Mass Measurement of the Decaying Bino at the LHC,''
Phys.\ Lett.\  B {\bf 672}, 339 (2009)
[arXiv:0807.4987 [hep-ph]].
%%CITATION = PHLTA,B672,339;%%




\bibitem{Gherghetta:1999sw}
T.~Gherghetta, G.~F.~Giudice and J.~D.~Wells,
%``Phenomenological consequences of supersymmetry with anomaly induced masses,''
Nucl.\ Phys.\ B {\bf 559}, 27 (1999)
[hep-ph/9904378];
%%CITATION = HEP-PH/9904378;%%
%\bibitem{Moroi:1999zb}
T.~Moroi and L.~Randall,
%``Wino cold dark matter from anomaly mediated SUSY breaking,''
Nucl.\ Phys.\ B {\bf 570}, 455 (2000)
[hep-ph/9906527].
%%CITATION = HEP-PH/9906527;%%

\bibitem{Hisano:2005ec}
J.~Hisano, S.~Matsumoto, O.~Saito and M.~Senami,
%``Heavy wino-like neutralino dark matter annihilation into antiparticles,''
Phys.\ Rev.\ D {\bf 73}, 055004 (2006)
[hep-ph/0511118];
%%CITATION = HEP-PH/0511118;%%
%\bibitem{Hisano:2006nn}
J.~Hisano, S.~Matsumoto, M.~Nagai, O.~Saito and M.~Senami,
%``Non-perturbative effect on thermal relic abundance of dark matter,''
Phys.\ Lett.\ B {\bf 646}, 34 (2007)
[hep-ph/0610249].
%%CITATION = HEP-PH/0610249;%%

\bibitem{Hisano:2010ct}
J.~Hisano, K.~Ishiwata and N.~Nagata,
%``Gluon contribution to the dark matter direct detection,''
Phys.\ Rev.\ D {\bf 82}, 115007 (2010)
[arXiv:1007.2601 [hep-ph]];
%%CITATION = ARXIV:1007.2601;%%
%\bibitem{Hisano:2011cs}
J.~Hisano, K.~Ishiwata, N.~Nagata and T.~Takesako,
%``Direct Detection of Electroweak-Interacting Dark Matter,''
JHEP {\bf 07}, 005 (2011)
[arXiv:1104.0228 [hep-ph]];
%%CITATION = ARXIV:1104.0228;%%
%\bibitem{Hill:2011be}
R.~J.~Hill and M.~P.~Solon,
%``Universal behavior in the scattering of heavy, weakly interacting dark matter on nuclear targets,''
Phys.\ Lett.\ B {\bf 707}, 539 (2012)
[arXiv:1111.0016 [hep-ph]];
%%CITATION = ARXIV:1111.0016;%%
%\cite{Hisano:2012wm}
%\bibitem{Hisano:2012wm} 
  J.~Hisano, K.~Ishiwata and N.~Nagata,
  %``Direct Search of Dark Matter in High-Scale Supersymmetry,''
  Phys.\ Rev.\ D {\bf 87}, 035020 (2013)
  [arXiv:1210.5985 [hep-ph]].
  %%CITATION = ARXIV:1210.5985;%%
  %9 citations counted in INSPIRE as of 24 Jul 2013



\bibitem{Hisano:2004ds}
J.~Hisano, S.~.Matsumoto, M.~M.~Nojiri and O.~Saito,
%``Non-perturbative effect on dark matter annihilation and gamma ray signature from galactic center,''
Phys.\ Rev.\ D {\bf 71}, 063528 (2005)
[hep-ph/0412403].
%%CITATION = HEP-PH/0412403;%%

%\cite{Moroi:2011ab}
\bibitem{Moroi:2011ab} 
  T.~Moroi and K.~Nakayama,
  %``Wino LSP detection in the light of recent Higgs searches at the LHC,''
  Phys.\ Lett.\ B {\bf 710}, 159 (2012)
  [arXiv:1112.3123 [hep-ph]].
  %%CITATION = ARXIV:1112.3123;%%

%\cite{Cohen:2013ama}
\bibitem{Cohen:2013ama} 
  T.~Cohen, M.~Lisanti, A.~Pierce and T.~R.~Slatyer,
  %``Wino Dark Matter Under Siege,''
  arXiv:1307.4082 [hep-ph].
  %%CITATION = ARXIV:1307.4082;%%
  %1 citations counted in INSPIRE as of 24 Jul 2013

%\cite{Fan:2013faa}
\bibitem{Fan:2013faa} 
  J.~Fan and M.~Reece,
  %``In Wino Veritas? Indirect Searches Shed Light on Neutralino Dark Matter,''
  arXiv:1307.4400 [hep-ph].
  %%CITATION = ARXIV:1307.4400;%%
  %1 citations counted in INSPIRE as of 24 Jul 2013





%\cite{Izawa:1997gs}
\bibitem{Izawa:1997gs} 
  K.~I.~Izawa, Y.~Nomura, K.~Tobe and T.~Yanagida,
  %``Direct transmission models of dynamical supersymmetry breaking,''
  Phys.\ Rev.\ D {\bf 56}, 2886 (1997)
  [hep-ph/9705228];
  %%CITATION = HEP-PH/9705228;%%
  %124 citations counted in INSPIRE as of 22 Jul 2013
%\cite{Nomura:1997uu}
\bibitem{Nomura:1997uu} 
  Y.~Nomura and K.~Tobe,
  %``Phenomenological aspects of a direct transmission model of dynamical supersymmetry breaking with the gravitino mass m(3/2) < 1-keV,''
  Phys.\ Rev.\ D {\bf 58}, 055002 (1998)
  [hep-ph/9708377].
  %%CITATION = HEP-PH/9708377;%%
  %37 citations counted in INSPIRE as of 22 Jul 2013

%\cite{Komargodski:2009jf}
\bibitem{Komargodski:2009jf} 
  Z.~Komargodski and D.~Shih,
  %``Notes on SUSY and R-Symmetry Breaking in Wess-Zumino Models,''
  JHEP {\bf 0904}, 093 (2009)
  [arXiv:0902.0030 [hep-th]].
  %%CITATION = ARXIV:0902.0030;%%
  %106 citations counted in INSPIRE as of 24 Jul 2013


%\cite{Shirai:2010rr}
\bibitem{Shirai:2010rr} 
  S.~Shirai, M.~Yamazaki and K.~Yonekura,
  %``Aspects of Non-minimal Gauge Mediation,''
  JHEP {\bf 1006}, 056 (2010)
  [arXiv:1003.3155 [hep-ph]].
  %%CITATION = ARXIV:1003.3155;%%
  %18 citations counted in INSPIRE as of 24 Jul 2013

%\cite{Ibe:2012dd}
\bibitem{Ibe:2012dd} 
  M.~Ibe and R.~Sato,
  %``A 125 GeV Higgs Boson Mass and Gravitino Dark Matter in R-invariant Direct Gauge Mediation,''
  Phys.\ Lett.\ B {\bf 717}, 197 (2012)
  [arXiv:1204.3499 [hep-ph]].
  %%CITATION = ARXIV:1204.3499;%%
  %3 citations counted in INSPIRE as of 24 Jul 2013








%\cite{Saito:2012bb}
\bibitem{Saito:2012bb} 
  R.~Saito and S.~Shirai,
  %``Gravitational Wave Probe of High Supersymmetry Breaking Scale,''
  Phys.\ Lett.\ B {\bf 713}, 237 (2012)
  [arXiv:1201.6589 [hep-ph]].
  %%CITATION = ARXIV:1201.6589;%%
  %7 citations counted in INSPIRE as of 30 May 2013




%\cite{Toharia:2005gm}
\bibitem{Toharia:2005gm} 
  M.~Toharia and J.~D.~Wells,
  %``Gluino decays with heavier scalar superpartners,''
  JHEP {\bf 0602}, 015 (2006)
  [hep-ph/0503175].
  %%CITATION = HEP-PH/0503175;%%



%\cite{Gambino:2005eh}
\bibitem{Gambino:2005eh} 
  P.~Gambino, G.~F.~Giudice and P.~Slavich,
  %``Gluino decays in split supersymmetry,''
  Nucl.\ Phys.\ B {\bf 726}, 35 (2005)
  [hep-ph/0506214].
  %%CITATION = HEP-PH/0506214;%%

%\cite{Sato:2012xf}
\bibitem{Sato:2012xf} 
  R.~Sato, S.~Shirai and K.~Tobioka,
  %``Gluino Decay as a Probe of High Scale Supersymmetry Breaking,''
  JHEP {\bf 1211}, 041 (2012)
  [arXiv:1207.3608 [hep-ph]].
  %%CITATION = ARXIV:1207.3608;%%
  %9 citations counted in INSPIRE as of 30 May 2013



%\cite{Arvanitaki:2005fa}
\bibitem{Arvanitaki:2005fa} 
  A.~Arvanitaki, C.~Davis, P.~W.~Graham, A.~Pierce and J.~G.~Wacker,
  %``Limits on split supersymmetry from gluino cosmology,''
  Phys.\ Rev.\ D {\bf 72}, 075011 (2005)
  [hep-ph/0504210].
  %%CITATION = HEP-PH/0504210;%%
  %72 citations counted in INSPIRE as of 05 Jul 2013

%\cite{Machacek:1981ic}
\bibitem{Machacek:1981ic} 
  M.~E.~Machacek and M.~T.~Vaughn,
  %``Fermion And Higgs Masses As Probes Of Unified Theories,''
  Phys.\ Lett.\ B {\bf 103}, 427 (1981).
  %%CITATION = PHLTA,B103,427;%%
  %66 citations counted in INSPIRE as of 21 Jun 2013



%\cite{Ibe:2012sx}
\bibitem{Ibe:2012sx} 
  M.~Ibe, S.~Matsumoto and R.~Sato,
  %``Mass Splitting between Charged and Neutral Winos at Two-Loop Level,''
  Phys.\ Lett.\ B {\bf 721}, 252 (2013)
  [arXiv:1212.5989 [hep-ph]].
  %%CITATION = ARXIV:1212.5989;%%
  %7 citations counted in INSPIRE as of 22 Jul 2013

\bibitem{winodirect}
ATLAS collaboration, ATLAS-CONF-2013-069



%\cite{Barbieri:1987ed}
\bibitem{Barbieri:1987ed} 
  R.~Barbieri, G.~Gamberini, G.~F.~Giudice and G.~Ridolfi,
  %``Constraining Supergravity Models From Gluino Production,''
  Nucl.\ Phys.\ B {\bf 301}, 15 (1988);
  %%CITATION = NUPHA,B301,15;%%

  
%\cite{Gabbiani:1996hi}
\bibitem{Gabbiani:1996hi} 
  F.~Gabbiani, E.~Gabrielli, A.~Masiero and L.~Silvestrini,
  %``A Complete analysis of FCNC and CP constraints in general SUSY extensions of the standard model,''
  Nucl.\ Phys.\ B {\bf 477}, 321 (1996)
  [hep-ph/9604387].
  %%CITATION = HEP-PH/9604387;%%
  %1043 citations counted in INSPIRE as of 06 Jun 2013
  
  
  
  
    
%\cite{Altmannshofer:2009ne}
\bibitem{Altmannshofer:2009ne} 
  W.~Altmannshofer, A.~J.~Buras, S.~Gori, P.~Paradisi and D.~M.~Straub,
  %``Anatomy and Phenomenology of FCNC and CPV Effects in SUSY Theories,''
  Nucl.\ Phys.\ B {\bf 830}, 17 (2010)
  [arXiv:0909.1333 [hep-ph]].
  %%CITATION = ARXIV:0909.1333;%%
  %147 citations counted in INSPIRE as of 06 Jun 2013
  
  
  
  
  
  %\cite{McKeen:2013dma}
\bibitem{McKeen:2013dma} 
  D.~McKeen, M.~Pospelov and A.~Ritz,
  %``EDM Signatures of PeV-scale Superpartners,''
  arXiv:1303.1172 [hep-ph].
  %%CITATION = ARXIV:1303.1172;%%
  %6 citations counted in INSPIRE as of 10 Jun 2013

  
  
  
  
%\cite{Bona:2007vi}
\bibitem{Bona:2007vi} 
  M.~Bona {\it et al.}  [UTfit Collaboration],
  %``Model-independent constraints on $\Delta$ F=2 operators and the scale of new physics,''
  JHEP {\bf 0803}, 049 (2008)
  [arXiv:0707.0636 [hep-ph]].
  %%CITATION = ARXIV:0707.0636;%%
  %254 citations counted in INSPIRE as of 21 Feb 2013
  
   %\cite{Bevan:2012waa}
\bibitem{Bevan:2012waa} 
  A.~J.~Bevan {\it et al.}  [UTfit Collaboration],
  %``The UTfit Collaboration Average of D Meson Mixing Data: Spring 2012,''
  JHEP {\bf 1210}, 068 (2012)
  [arXiv:1206.6245 [hep-ph]].
  %%CITATION = ARXIV:1206.6245;%%
  %4 citations counted in INSPIRE as of 10 Jun 2013
 
  
  
\bibitem{UTfit} 
\url{http://www.utfit.org/UTfit/ResultsWinter2013PreMoriond}
  

 
%\cite{Hisano:2004tf}
\bibitem{Hisano:2004tf} 
  J.~Hisano and Y.~Shimizu,
  %``Hadronic EDMs induced by the strangeness and constraints on supersymmetric CP phases,''
  Phys.\ Rev.\ D {\bf 70}, 093001 (2004)
  [hep-ph/0406091].
  %%CITATION = HEP-PH/0406091;%%
  %76 citations counted in INSPIRE as of 10 Jun 2013
 
%\cite{Baker:2006ts}
\bibitem{Baker:2006ts} 
  C.~A.~Baker, D.~D.~Doyle, P.~Geltenbort, K.~Green, M.~G.~D.~van der Grinten, P.~G.~Harris, P.~Iaydjiev and S.~N.~Ivanov {\it et al.},
  %``An Improved experimental limit on the electric dipole moment of the neutron,''
  Phys.\ Rev.\ Lett.\  {\bf 97}, 131801 (2006)
  [hep-ex/0602020].
  %%CITATION = HEP-EX/0602020;%%
  %395 citations counted in INSPIRE as of 11 Jun 2013
 
 
%\cite{Griffith:2009zz}
\bibitem{Griffith:2009zz} 
  W.~C.~Griffith, M.~D.~Swallows, T.~H.~Loftus, M.~V.~Romalis, B.~R.~Heckel and E.~N.~Fortson,
  %``Improved Limit on the Permanent Electric Dipole Moment of Hg-199,''
  Phys.\ Rev.\ Lett.\  {\bf 102}, 101601 (2009).
  %%CITATION = PRLTA,102,101601;%%
  %110 citations counted in INSPIRE as of 11 Jun 2013
  
  
%\cite{Hewett:2004nw}
\bibitem{Hewett:2004nw} 
  J.~L.~Hewett, B.~Lillie, M.~Masip and T.~G.~Rizzo,
  %``Signatures of long-lived gluinos in split supersymmetry,''
  JHEP {\bf 0409}, 070 (2004)
  [hep-ph/0408248].
  %%CITATION = HEP-PH/0408248;%%
  %107 citations counted in INSPIRE as of 17 Jul 2013


  
  %\cite{Cheung:2004ad}
\bibitem{Cheung:2004ad} 
  K.~Cheung and W.~-Y.~Keung,
  %``Split supersymmetry, stable gluino, and gluinonium,''
  Phys.\ Rev.\ D {\bf 71}, 015015 (2005)
  [hep-ph/0408335].
  %%CITATION = HEP-PH/0408335;%%
  %62 citations counted in INSPIRE as of 17 Jul 2013

  
  

%\cite{Bartl:2011wq}
\bibitem{Bartl:2011wq} 
  A.~Bartl, H.~Eberl, E.~Ginina, B.~Herrmann, K.~Hidaka, W.~Majerotto and W.~Porod,
  %``Flavour violating gluino three-body decays at LHC,''
  Phys.\ Rev.\ D {\bf 84}, 115026 (2011)
  [arXiv:1107.2775 [hep-ph]].
  %%CITATION = ARXIV:1107.2775;%%
  %4 citations counted in INSPIRE as of 30 May 2013





%\cite{Arvanitaki:2005nq}
\bibitem{Arvanitaki:2005nq} 
  A.~Arvanitaki, S.~Dimopoulos, A.~Pierce, S.~Rajendran and J.~G.~Wacker,
  %``Stopping gluinos,''
  Phys.\ Rev.\ D {\bf 76}, 055007 (2007)
  [hep-ph/0506242].
  %%CITATION = HEP-PH/0506242;%%
  %87 citations counted in INSPIRE as of 11 Jun 2013





%\cite{Ito:2011xs}
\bibitem{Ito:2011xs} 
  T.~Ito, K.~Nakaji and S.~Shirai,
  %``Identifying the Origin of Longevity of Metastable Stau at the LHC,''
  Phys.\ Lett.\ B {\bf 706}, 314 (2012)
  [arXiv:1104.2101 [hep-ph]].
  %%CITATION = ARXIV:1104.2101;%%
  %3 citations counted in INSPIRE as of 30 May 2013

\bibitem{fv_in_gluino}
  R. Sato, S. Shirai and K. Tobioka, in preparation.


%\cite{Moroi:2013sfa}
\bibitem{Moroi:2013sfa} 
  T.~Moroi and M.~Nagai,
  %``Probing Supersymmetric Model with Heavy Sfermions Using Leptonic Flavor and CP Violations,''
  Phys.\ Lett.\ B {\bf 723}, 107 (2013)
  [arXiv:1303.0668 [hep-ph]],
  %%CITATION = ARXIV:1303.0668;%%
  %5 citations counted in INSPIRE as of 19 Jun 2013
%\cite{Moroi:2013vya}
%\bibitem{Moroi:2013vya} 
  T.~Moroi, M.~Nagai and T.~T.~Yanagida,
  %``Lepton Flavor Violations in High-Scale SUSY with Right-Handed Neutrinos,''
  arXiv:1305.7357 [hep-ph].
  %%CITATION = ARXIV:1305.7357;%%






\end{thebibliography}
\end{document}